\documentclass[useAMS,usenatbib,a4paper,fleqn]{mn2e}

\usepackage{amsfonts,amssymb,amsmath}
\usepackage{aas_macros}
\usepackage{times,varioref,multirow,textcomp,comment}
\usepackage[usenames,dvipsnames,svgnames,hyperref]{xcolor}
\usepackage[pdftex]{graphicx}
\usepackage[
    pdftex,
    a4paper=true,
    plainpages=true,
    pdfpagelabels,
    breaklinks=true,
    bookmarks=true,
    bookmarksopen=false,
    bookmarksopenlevel=2
    bookmarksnumbered=true,
    bookmarkstype=toc,
    colorlinks=true,
    citecolor=RoyalBlue,
    linkcolor=ForestGreen,
    menucolor=Teal,
    urlcolor=DarkOrange,
]{hyperref}

\providecommand{\adsurl}[1]{\href{#1}{ADS}}

\hypersetup{
  pdfauthor = {Pascal Jahan Elahi},
  pdfkeywords = {Code, Numerical},
  pdftitle = {Streams Going Notts}
}

\def \merit{\mathcal{M}}
\def \purity{\mathcal{P}_{\rm tot}}
\def \recoveryfrac{\mathcal{R}}
\def \Tsub{T_{\rm sub}}
\def \Tsubtid{T^{\rm sub}_{\rm tid}}
\def \Ttid{T_{\rm tid}}

\def \fracE{f_E}

\def \Vr{\mathcal{V}_{\rm r}}
\def \Thetaop{\Theta_{\rm op}}
\def \ellx{\ell_{\rm x}}

\def \ELL{\mathcal{L}}
\def \stf{{\sc stf}}
\def \velociraptor{{\sc veloci}raptor}

\def \FOF{{\sc fof}}
\def \6DFOF{{\sc 6dfof}}

\def \rockstar{{\sc rockstar}}
\def \hot6d{{\sc hot6d}}
\def \hst{{\sc s-tracker}}
\def \hbt{{\sc hbt}}
\def \Gadget2{{\sc gadget-2}}

\newcommand{\Eqref}[1]{Eq.~(\ref{#1})}
\newcommand{\Figref}[1]{Fig.~\ref{#1}}
\newcommand{\Secref}[1]{\S\ref{#1}}  
\newcommand{\Tableref}[1]{Table~\ref{#1}}

\defcitealias{nfw}{NFW}
\defcitealias{enbid}{{\sc enbid}}
\defcitealias{skid}{{\sc skid}}
\defcitealias{enlink}{{\sc enlink}}
\defcitealias{subfind}{{\sc subfind}}
\defcitealias{hsf}{{\sc hsf}}

\begin{document}
\title[Streams Going Notts]{Streams Going Notts: The tidal debris finder comparison project}
\author[P.J.~Elahi, {\it et al}.] {
  Pascal~J.~Elahi$^{1,2,3}$\thanks{\href{mailto:pelahi@shao.ac.cn}{pelahi@shao.ac.cn}},
  Jiaxin~Han$^{2,4}$,
  Hanni~Lux$^{3,9}$,
  Yago Ascasibar$^{5}$,
  Peter~Behroozi$^{6,7,8}$,
  \newauthor
  Alexander Knebe$^5$,
  Stuart I. Muldrew$^3$, 
  Julian Onions$^3$,
  Frazer Pearce$^3$
\\
$^{1}$  Sydney Institute for Astronomy, University of Sydney, Sydney, NSW 2006, Australia \\
$^{2}$	Key Laboratory for Research in Galaxies and Cosmology, Shanghai Astronomical Observatory, Shanghai, 200030, China \\
$^{3}$	School of Physics \& Astronomy, University of Nottingham, Nottingham, NG7 2RD, UK\\
$^{4}$  Institute of Computational Cosmology, Department of Physics, University of Durham, Science Laboratories, South Road, Durham DH1 3LE, UK\\
$^{5}$	Departamento de F\'{i}sica Te\'{o}rica, M\'{o}dulo C-15, Facultad de Ciencias, Universidad Aut\'{o}noma de Madrid, 28049 Cantoblanco, Madrid, Spain\\
$^{6}$	Kavli Institute for Particle Astrophysics and Cosmology, Stanford, CA 94309, USA\\
$^{7}$	Physics Department, Stanford University, Stanford, CA 94305, USA\\
$^{8}$	SLAC National Accelerator Laboratory, Menlo Park, CA 94025, USA\\
$^{9}$  Department of Physics, University of Oxford, Denys Wilkinson Building, Keble Road, Oxford, OX1 3RH, UK\\
}
\maketitle

\pdfbookmark[1]{Abstract}{sec:abstract}
\begin{abstract}
While various codes exist to systematically and robustly find haloes and subhaloes in cosmological simulations \cite[][]{knebe2011,onions2012}, this is the first work to introduce and rigorously test codes that find tidal debris (streams and other unbound substructure) in fully cosmological simulations of structure formation. We use one tracking and three non-tracking codes to identify substructure (bound and unbound) in a Milky Way type simulation from the Aquarius suite \cite[][]{springel2008} and post-process their output with a common pipeline to determine the properties of these substructures in a uniform way. By using output from a fully cosmological simulation, we also take a step beyond previous studies of tidal debris that have used simple toy models. We find that both tracking and non-tracking codes agree well on the identification of subhaloes and more importantly, the {\em unbound tidal features} associated with them. The distributions of basic properties of the total substructure distribution (mass, velocity dispersion, position) are recovered with a scatter of $\sim20\%$. Using the tracking code as our reference, we show that the non-tracking codes identify complex tidal debris with purities of $\sim40\%$. Analysing the results of the substructure finders, we find that the general distribution of {\em substructures} differ significantly from the distribution of bound {\em subhaloes}. Most importantly, both bound and unbound {\em substructures} together constitute $\sim18\%$ of the host halo mass, which is a factor of $\sim2$ higher than the fraction in self-bound {\em subhaloes}. However, this result is restricted by the remaining challenge to cleanly define when an unbound structure has become part of the host halo. Nevertheless, the more general substructure distribution provides a more complete picture of a halo's accretion history.
\end{abstract}
\maketitle

\section{Introduction}\label{sec:intro}
A series of papers have been devoted to comparing the results of various (sub)halo finders on the same data sets \citep{knebe2011,onions2012,knebe2012a,knebe2013a}. As a first paper of its kind, this work introduces and compares systematic ways of finding tidal debris from disrupted subhaloes in {\em fully cosmological simulations}. The motivation for this study is quite simple: a wealth of information has been extracted from studies of subhaloes in cosmological simulations \cite[e.g][]{moore1999,gao2004,gill2005,springel2008,diemand2008,stadel2008,elahi2009,libeskind2010,knebe2011b,gao2012}. The growing need for reliable and fast analysis tools of N-body simulations has led to a surge in the number of (sub)halo finders available. Subhaloes, however, are not the only substructures present in haloes. A halo's tidal field will distort and destroy subhaloes. These disrupted substructures leave behind dynamically distinct tidal streams and shells, which may constitute a substantial fraction of the host's mass (e.g.~\citealp{helmi2003,knebe2005,warnick2008,cooper2010,libeskind2011,maciejewski2011}). 

\par
The properties and distribution of tidal debris from fully cosmological simulations offers a new window on the formation history of haloes. For instance, streams offer a way of sampling the underlying potential and possibly testing different theories of gravity (e.g.~\citealp{johnston2005,law2010,penarrubia2012}), although studies have typically focused on the evolution of disrupted satellites in idealised situations. Deep surveys, such as GAIA \citep{gaia2008,gomez2010}, will allow observers to search for stellar streams, providing a new window with which to test the current paradigm of structure formation or 'Near-Field' cosmology \citep{freeman2002}. Finally, the number, velocity profile and mass of these streams will have important ramifications for direct and indirect dark matter searches (e.g.~\citealp{stiff2001,vogelsberger2008,fairbairn2009,kuhlen2010,alhen2010,cdms2011}). 

\par
Tidally disrupted substructures present a special challenge for many substructure finders, as most identify substructures by searching for clustering in configuration-space. Tidal debris need not show up as an overdensity, consequently configuration-space finders are completely ill-suited to this task. In the past, the inability of these algorithms to detect tidal streams has mainly been addressed by tracking previously tagged particles of an accreted halo in time (e.g.~\citealp{bullock2005,warnick2008,cooper2010,helmi2011,rashkov2012,kuhlen2012}). However, requiring temporal information severely limits the search for streams, and such techniques cannot be directly applied to observational data sets, which only provide an instantaneous snapshot of ``particle'' (star) positions. Observational missions such as GAIA hope to measure the phase-space positions of millions of stars \citep{gomez2010,lindegren2012}. Searching through this data set for kinematic structures, such as stellar streams, in our Galaxy will require tools that can robustly identify these structures {\em without} the need to follow their evolution. 

\par
Furthermore, simply tracking particles tagged by a structure finder is not enough. An accreted substructure will begin to phase-mix as it orbits the host. After many orbital periods in an evolving potential, it is unlikely that a tidally disrupted substructure will still occupy a well-defined volume in phase-space, particularly since N-body codes {\em do not} explicitly conserve phase-space density. Additionally, an accreted halo does not stop growing simply because it has entered the virial radius of another host halo. In the low density environment at the outskirts of a host halo, these new subhaloes may accrete some mass (see for instance \citealp{hbt}). Therefore, identifying coherent dynamical structures such as unbound tidal debris by a tracking algorithm necessitates the application of a dynamically motivated criterion when pruning particles. For subhaloes this takes the form of removing unbound particles. For tidal debris such as streams, one can make use of the fact that they are kinematically cold, physically extended substructures that are tightly distributed about integrals of motion in slowly varying potentials (e.g.~\mbox{\citealp{helmi1999,penarrubia2006}};\\\mbox{\citealp{mcmillan2008}}). 

\par
However, one does not have to solely rely on tracking particles to identify streams. As streams should occupy specific volumes in phase-space and appear clustered in orbit space, algorithms which use velocity information should in principle be able to identify tidal debris (e.g.~\citealp{enlink,maciejewski2011,ascasibar2010,rockstar,elahi2011}). One of the goals of this work is to examine in a systematic fashion how well these non-tracking codes recover tidal streams by comparing them to a tracking method using {\em fully cosmological simulations}. To this end, we cross-correlate the substructures identified and examine the recovered properties of these substructures. We note that the performance of some of the codes mentioned have been tested using non-cosmological simulations, naively assuming that toy models of satellite disruption captures the full complexity of structure formation. Here we take a far more realistic approach and asses the ability of codes to recover tidal debris in fully cosmological simulations, where tidal debris can be heated by encounters with other substructures and can originate from (sub)haloes which contain their own (sub)substructure. These tools can, in principle, be applied to observational data to search for stellar structures. However, here we focus only on dark matter structures, which occupy larger phase-space volumes than stars and hence phase-mix quicker than stellar structures and are consequently harder to detect.

\par
Our paper is organized as follows: in section \ref{sec:data}, we briefly summarize the data sets used and outline the common post-processing pipeline. We summarize the different algorithms used in \Secref{sec:algorithms}.  We present a detailed comparison in \Secref{sec:results}. First, in \Secref{sec:classification} we present several substructure properties used to determine whether a substructure is a subhalo or should be considered tidal debris. We give an overview of the general results and then present how well different types of substructures are recovered, focusing specifically on unbound substructures in \Secref{sec:generalresults}-\ref{sec:cross-matching}. The properties of the substructure distribution function are discussed in \Secref{sec:substructureprop}. We summarize our results in \Secref{sec:discussion}, and conclude by outlining future studies and improvements in the search for and study of tidal debris in \Secref{sec:conclusion}.

\section{Numerical Methods}\label{sec:data}
\subsection{Simulations}\label{sec:sims}
The data used in most of this comparison study comes from the Aquarius project \citep{springel2008}. It consists of multiple dark matter only re-simulations of a Milky Way-like halo at a variety of resolutions performed using {\sc gadget3} (based on {\sc gadget2}, \citealp{gadget2}). We have used in the main the Aquarius-A halo dataset at $z = 0$ for this project. The underlying cosmology for the Aquarius simulations is
\begin{gather*}
 \Omega_m=0.25,\ \Omega_\Lambda=0.75,\ \sigma_8=0.9,\ n_s=1,\ h=0.73
\end{gather*}
consistent with that used in \cite{springel2005} and only mildly inconsistent with the latest WMAP data \citep{wmap9}. Precise details on the setup and performance of these simulations can be found in \cite{springel2008}. Here we focus on the level 4 resolution halo which is composed of $6\times10^{6}$ particles.

\subsection{Post-processing pipeline}\label{sec:pipeline}
As with the other studies in this series, all substructure finder results are subjected to the same post-processing pipeline. Each code representative was asked to return a list of substructures and the list of particle ID's belong to that substructure. Note that two of the finders used in this comparison, \rockstar\ and \hot6d, allow particles to belong to multiple substructures in the substructure hierarchy. For example, a particle residing within a subsubhalo is included in the list of particles for the subsubhalo, its host subhalo, and the halo itself. The two others only allow a particle to be associated with one substructure at the deepest level of the substructure hierarchy, which in the example given would be the subsubhalo composed on the fewest particles. Either of these approaches are acceptable. For the sake of clarity we prune the lists so that particles only belong to the substructure composed of the fewest number of particles. 

\par
We then calculate various properties for each substructure to quantify its dynamical state and phase-space morphology. We iterate to find a substructure's centre-of-mass till the radius encloses the inner most, densest $10\%$ of the particles. It should be emphasized that for completely disrupted substructures, such a quantity does not have the same physical significance. The centre-of-mass velocity, ${\bf V}_{\rm cm}$, is determined using the same region as the centre-of-mass.

\par
To determine the dynamical state of a substructure we use it's energy distribution. The specific energy of the $i^{\rm th}$ particle is
\begin{align}
    E_i=\phi_i+\tfrac{1}{2}\left({\bf v}_i-{\bf V}_{\rm cm}\right)^2,\label{eqn:energy}
\end{align}
where the potential is calculated only using the particles in the group. Studies of (sub)haloes often require that the substructure be ``self-bound'', that is all particles must have $E_i<0$. Tidal debris, by its very nature, is not self-bound. Thus, we quantify the dynamical state of a substructure using the fraction of particles with negative energy, $\fracE$. 

\par
Next we determine how anisotropic the substructure is in both configuration- and velocity-space. To determine the physical morphology, we follow \cite{dubinski1991} and \cite{allgood2006} and diagonalize the weighted moments of the reduced inertia tensor
\begin{equation}
    \tilde{I}_{i,j}=\sum\limits_n \frac{m_n x^\prime_{i,n} x^\prime_{j,n}}{(r^\prime_{n})^2}.\label{eqn:inertia}
\end{equation}
The ellipsoidal distance between the subhalo's centre of mass and the $n$th particle is
\begin{equation}
    (r^\prime_n)^2=(x^\prime_n)^2+(y^\prime_n/q)^2+(z^\prime_n/s)^2.
\end{equation}
Here $q^2=\lambda_2/\lambda_1$ \& $s^2=\lambda_3/\lambda_1$ are the intermediate and minor axis ratios respectively, $\lambda_i$ are the ordered eigenvalues of $\tilde{I}_{i,j}$, and ${\bf x}^\prime$ coordinates are in the eigenvector frame.

\par
This method accurately recovers the orientation of an ellipsoid. However, due to the inverse radial weighting, the axis ratios tend to be systematically overestimated, making structures appear slightly more spherical than they really are \citep{zemp2011}. The reason we use this tensor in spite of this bias is to ensure that the morphology is not severely skewed by the presence of a few very distant particles. We emphasize that the physical meaning of these eigenvalues for completely disrupted amorphous substructures is not obvious, unlike the case for ellipsoidal substructures. Tidal debris may be spread over several orbits with regions containing particles that are at different orbital phases, contain kinks and radial shells. However, most stream-like substructures will tend to have a small physical extent in directions normal to their orbital plane, i.e.~$s$ is small. Visually, we find stream like structures have $s\lesssim0.3$ (see \Figref{fig:phasespacemorph}). We emphasize that we do not use the inertia tensor to determine a substructure's {\em shape} but the deviation from a spherical distribution. Determining the shape of tidal debris, by using some combination of the inertia tensor and Minkowski functionals, is beyond the scope of this paper.

\par
We characterize the velocity anisotropy in a similar fashion using $q_v\equiv\sigma_2/\sigma_1$ and $s_v\equiv\sigma_3/\sigma_1$, where $\sigma^2_1,\ \sigma^2_2$ \& $\sigma^2_3$, are the ordered eigenvalues of the global velocity dispersion tensor, 
\begin{align}
    \sigma^2_{i,j}({\bf v})=\frac{1}{m_{\rm tot}}\sum\limits_n m_n (v_{i,n}-V_{i,{\rm cm}})(v_{j,n}-V_{j,{\rm cm}}). \label{eqn:veldisp}
\end{align}
For well virialized ellipsoidal substructures, $\sigma_1\sim\sigma_2\sim\sigma_3$ (see for instance \citealp{vogelsberger2009b} which shows that the velocity dispersion along the three main velocity eigenvectors of a halo do not differ significantly, though the halo's velocity distribution is still anisotropic). This tensor suffers from the same limitations as the inertia tensor. The physical interpretation of the eigenvalues for an ellipsoidal velocity distribution is clear whereas tidal debris need not be well characterized by a global velocity dispersion tensor. Again, as we do not attempt to characterize the velocity distribution of an entire substructure, this limitation is not an issue. We simply use the eigenvalues to quantify the anisotropy of the velocity distribution. As a substructure is stretched along its orbit, the inferred global dispersion along the direction of the orbit will appear to increase due to orbital gradients. In comparison, dispersions in directions orthogonal to the orbit should not evolve significantly \citep{binneytremaine2008,helmi1999,johnston2001}. In fact, if phase-space density is conserved, the structure should become kinematically colder as it is stretched physically. However, phase-space density is not explicitly conserved in N-body simulations, though in N-body simulations with sufficient resolution it is conserved \cite[][]{helmi2003}. Moreover, the presence of other substructures will tidally heat the substructure and substructure will not be on a closed orbit in the evolving triaxial potential of the host halo. Regardless, we should expect substructures to appear comparatively cold in one direction, i.e. small $s_v$.

\section{The Substructure Finders}\label{sec:algorithms}
Identifying dynamically distinct tidal debris in cosmological N-body simulations poses many intriguing challenges. Despite the wealth of algorithms designed to identify haloes \citep{knebe2011} and subhaloes \citep{onions2012}, many of these codes cannot be adapted to identify tidally disrupted substructures due to their need for an unbinding process \cite[see][]{onions2012b}. Here we briefly outline the substructure finders used in this study which can detect tidal debris, and clearly state the differences in the methodologies used. We should emphasize that these codes {\em do not} pass their candidate substructures through a so-called ``unbinding'' routine to prune particle lists, as is often done for {\em subhalo} finders and these codes have been run ``blind'', using parameters the code authors deem reasonable. For readers who are not interested in the details of any particular finder, we summarize the key differences in \Secref{sec:diffalgo}.

\subsection{S-Tracker}\label{sec:hst}
\hst\ is a tracing algorithm we have developed based on the Hierarchical Bound Tracing algorithm (\hbt) \citep{hbt}. The algorithm works in the time domain to trace the evolution of all FOF haloes accreted by other larger FOF haloes. As the algorithm tracks particles, the substructures identified originate from a distinct FOF halo and therefore no particle is incorrectly associated to a substructure. The key difference between this code and \hbt\ lies in the criterion used to prune particles in the substructure. \hbt\ was concerned with correctly identifying the descendant self-bound subhaloes, thus only retained particles with negative energy $E_i=\phi_i+K_i<0$. One might naively assume that relaxing the energy criterion, e.g. by requiring $\alpha\phi+K_i<0$ with $\alpha>1$, would be sufficient to recover unbound tidally disrupted substructures. However, this is not the case. As energy is measured relative to a reference frame, in this case the centre-of-mass velocity, different portions of a tidally disrupted object will have different relative energies. Setting a energy limit amounts to imposing a spherical velocity dispersion cut relative to the centre-of-mass velocity. Determining the optimal cut for any given tidally disrupted structure is not trivial. Consequently, such a criterion will never recover dynamically distinct shells or regions of a very physically elongated stream that lie a great distance from the centre-of-mass (for more discussions on unbinding see \citealp{knebe2013a}). 

\par
However, if particles originating from the same progenitor and initially occupying the same region in phase-space should remain neighbours in phase-space. We therefore track particles originating from the same progenitor that remain linked using a similar criterion to the one used in \cite{elahi2011}, that is 
\begin{gather}
    \frac{({\bf x}_i-{\bf x}_j)^2}{\ellx^2}<1,\notag\\
    1/\Vr\leq  v_i/v_j\leq \Vr,\notag\\
    \cos\Thetaop\leq \frac{{\bf v}_i\cdot{\bf v}_j}{v_i v_j}, \label{eqn:stffof}
\end{gather}
where $\ellx$ is the physical linking length, $V_r$ is the velocity ratio, and $\cos\Thetaop$ is the cosine between the two particles velocities. The first criterion is the standard \FOF\ criterion, linking particles that are separated by a distance less than $\ellx$ times the inter-particle spacing. The last two criteria ensure that particles have similar velocities. In this analysis we use $\ellx=0.35$\footnote{Note that haloes are found using $\ellx=0.2$.}, $\Vr=3$, and $\cos\Thetaop=0.85$. The algorithm also keeps track of substructure hierarchy. The hierarchy is ordered in accretion time akin to the hierarchy in a halo merger tree, that is if a halo is accreted by a larger halo, which is itself accreted at a later time, said halo resides at the 3$^{\rm rd}$ level in the substructure hierarchy. Consequently, if a particle is removed from one (sub)substructure, it is added to the pool of the parent (sub)structure candidate particles. Thus, in this approach substructures can grow via the {\em complete} disruption of internal substructure. 

\par
The algorithm requires that the time domain be well sampled, roughly 50 or more snapshots are needed \cite[see][for a discussion of the required temporal resolution, which depends on the merger rate and cosmology]{hbt}.  We use 128 snapshots, ensuring few substructures composed of $\lesssim50$ particles are not tracked by the algorithm. As a consequence of the zero contamination and low loss, we use the output of this code as our reference. However, it should be noted that since this algorithm must process every output, it is by far the slowest algorithm of the four studied here. 

\subsection{VELOCIraptor (a.k.a {\sc STF})}\label{sec:stf}
\velociraptor\ (formerly known as the STructure Finder, \stf) \citep{elahi2011} identifies substructures by utilising the fact that dynamically distinct substructures in a halo will have a local velocity distribution that differs significantly from the mean, i.e. smooth background halo. This method consists of two main steps, identifying particles that appear dynamically distinct and linking this outlier population using a Friends-of-Friends-like (FOF) approach. Specifically, outlier particles are identifying by estimating the mean velocity distribution function using a coarse grain approach and comparing the predicted distribution to that of a particle's local velocity distribution, which is calculated using a near-neighbour kernel technique. Specifically, the local velocity density is estimated using 32 nearest neighbours in velocity space drawn from a larger sample of 256 nearest neighbours in physical space. By taking the ratio of the local velocity distribution density relative to the expected mean velocity distribution density at a particle's phase-space position, the contrast of particles resident in substructure relative to those in the background are greatly enhanced. The scatter in this estimator is determined by examining the distribution of this ratio, $\mathcal{L}$, which is characterised by a Gaussian distribution corresponding to the virialized background, and numerous secondary peaks located at large values of the ratio arising from particles resident in substructure. The variance about the central main peak is estimated and only outlier particles, which have ratios lying several $n_{\ELL}\sigma$ away from the main peak, are searched. In this way, we quantify how dynamically different a particle is and the likelihood that a particle is resident in substructure. Particles are linked together if they satisfy the criteria outlined in \Eqref{eqn:stffof}. Initial candidates are found using an initial linking length, $\ell_{x,o}$ and then allowed to grow using a much larger linking length. In this analysis, we use $n_{\ELL}=2.5$, $\ell_{x,o}=0.10$, $\ellx=0.45$, $\Vr=2$, $\cos\Thetaop=0.95$. The algorithm also determines if a substructure is significant if the substructure's average ratio is greater than the expected value from Poisson fluctuations.

\subsection{ROCKSTAR}\label{sec:rockstar}
\rockstar\ (Robust Overdensity Calculation using K-Space Topologically Adaptive Refinement) is a phase-space halo finder designed to maximize halo consistency across timesteps \citep{rockstar}. The algorithm first selects particle groups with a 3D FOF variant with a very large linking length $(\ellx=0.28)$. For each main 3D FOF group, it then builds a hierarchy of 6D FOF subgroups in phase-space by progressively and adaptively reducing the linking lengths in configuration and velocity space, so that a tunable fraction ($70\%$, for this analysis) of particles are captured at each subgroup as compared to the immediate parent group. The distance between two particles is 
\begin{align}
  d(p_i, p_j) = \left(\frac{({\bf x}_i-{\bf x}_j)^2}{\sigma_x^2}+\frac{({\bf v}_i-{\bf v}_j)^2}{\sigma_v^2}\right)^{1/2},
\end{align}
where the configuration-space and velocity-space lengths defining the metric are proportional to the position and velocity dispersions, respectively, of the particles in the subgroup. Particles are linked if these distances are less than unity. When this is complete, \rockstar\ converts these 6DFOF subgroups into seed haloes beginning at the deepest level of the hierarchy. If a subgroup only contains one seed group, all particles in it are assigned to that group. Otherwise, when a subgroup contains multiple seeds, particles are assigned based on their proximity to seeds in phase-space using the coarse grain metric determined for that level in the tree. This process is repeated at all levels of the hierarchy until all particles in the base FOF group have been assigned to either a substructure or to the main halo. 

\subsection{HOT6D}\label{sec:hot6d}
\hot6d\ is a structure finder based on HOT+FiEstAS, which is a general-purpose clustering analysis tool still under development. This algorithm performs the unsupervised classification of a multidimensional data set by computing its Hierarchical Overdensity Tree (HOT), analogous to the Minimal Spanning Tree (MST) in Euclidean spaces, based on the density field returned by the Field Estimator for Arbitrary Spaces \citep{fiestas,ascasibar2010}. As explained in \cite{knebe2011} in the context of halo finding, \hot6d identifies substructures with density maxima in phase-space of particle positions and velocities. The boundary of a substructure is set by the isodensity contour crossing a saddle point, and its centre is defined as the density-weighted average of its constituent particles. 


\subsection{Key differences}\label{sec:diffalgo}
Although all the finders here effectively define substructures as regions clustered in phase-space, there are differences in the details inbuilt in the approach taken by each algorithm. 
\begin{itemize}
  \item \hst\ is a tracking code that applies phase-space criteria to a collection of particles originating from a distinct FOF-halo in order to prune the particle list and return substructure which occupies a specific region in phase-space. 
  \item \velociraptor\ was designed to identify physically underdense streams. It estimates the local velocity density distribution relative to the expected mean velocity distribution to identify candidate particles and then links them using a phase-space criterion. 
  \item \rockstar\ was designed to identify (sub)haloes, however, its use of phase-space information means that, in principle, it is capable of identifying tidally disrupted substructures. The algorithm uses a coarse grain phase-space metric and an adaptive 6D-FOF algorithm to link particles together.
  \item \hot6d was designed to identify clustering in arbitrary spaces. In the current version used here, it estimates the local phase-space density and uses density thresholds and saddle points to identify structures. 
\end{itemize}
We see that \hst\ explicitly assumes substructures originate from FOF-haloes, whereas the others do not. Both \hst\ and \velociraptor\ assume particles residing in a substructure will be physically close and on similar orbits. However the former prunes particles from a candidate substructure using phase-space criteria whereas the latter links particles together to identify candidates. Consequently, \hst\ uses more relaxed criteria when pruning particles. \rockstar\ and \hot6d\ weight configuration-space and velocity-space approximately equally whereas \velociraptor\ is biased towards clustering in velocity-space. 

\par
Therefore, some of the differences seen in the following sections are a result of different definitions\footnote{Interested readers are referred to \cite{knebe2013a} for a related discussion on the scatter in {\em subhaloes} arising from the various definitions of what constitutes a subhalo.}.

\section{How do algorithms compare?}\label{sec:results}
\subsection{Identifying tidal debris}\label{sec:classification}
\begin{figure}
    \centering
    \includegraphics[width=0.49\textwidth]{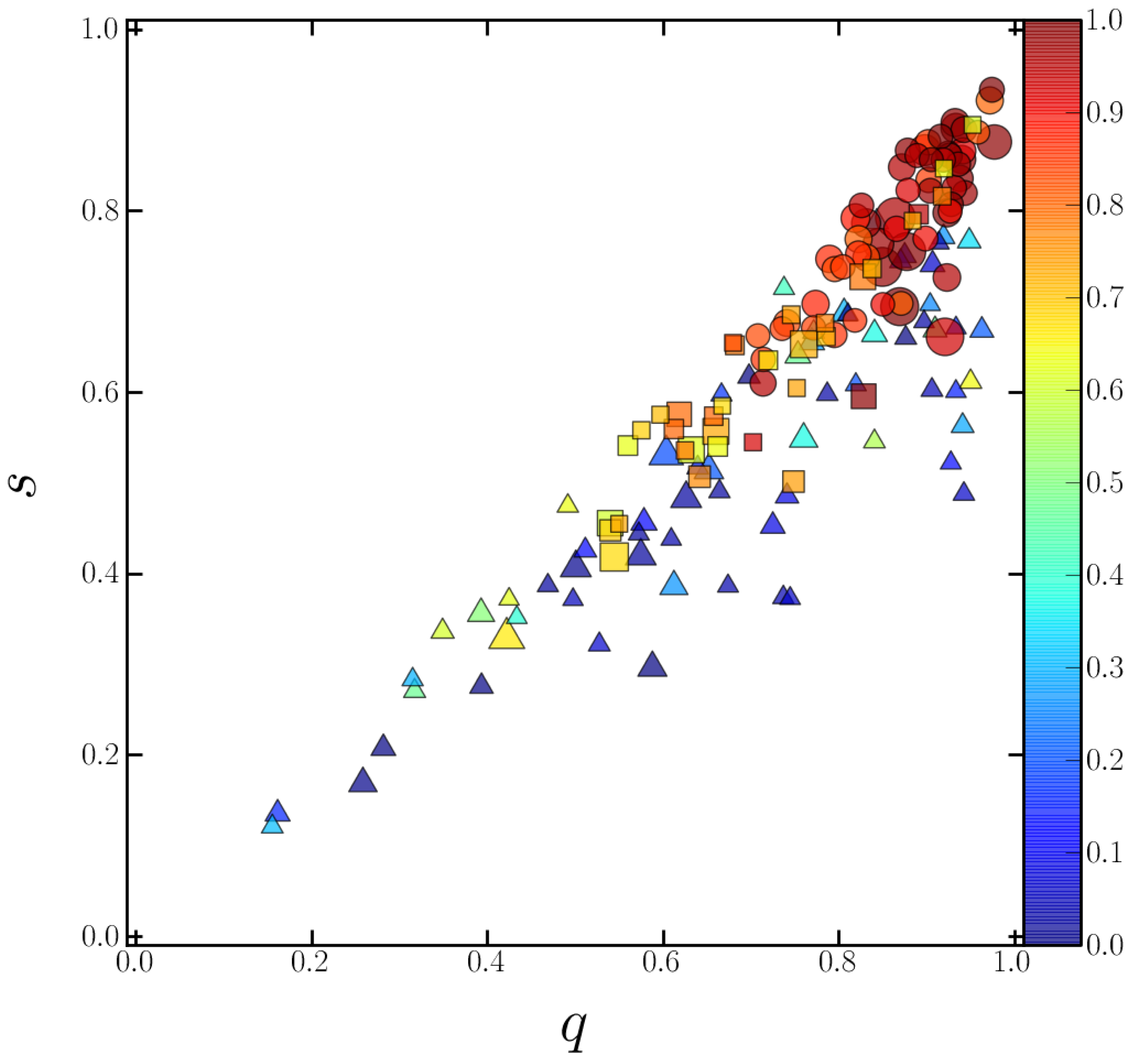}
    \includegraphics[width=0.49\textwidth]{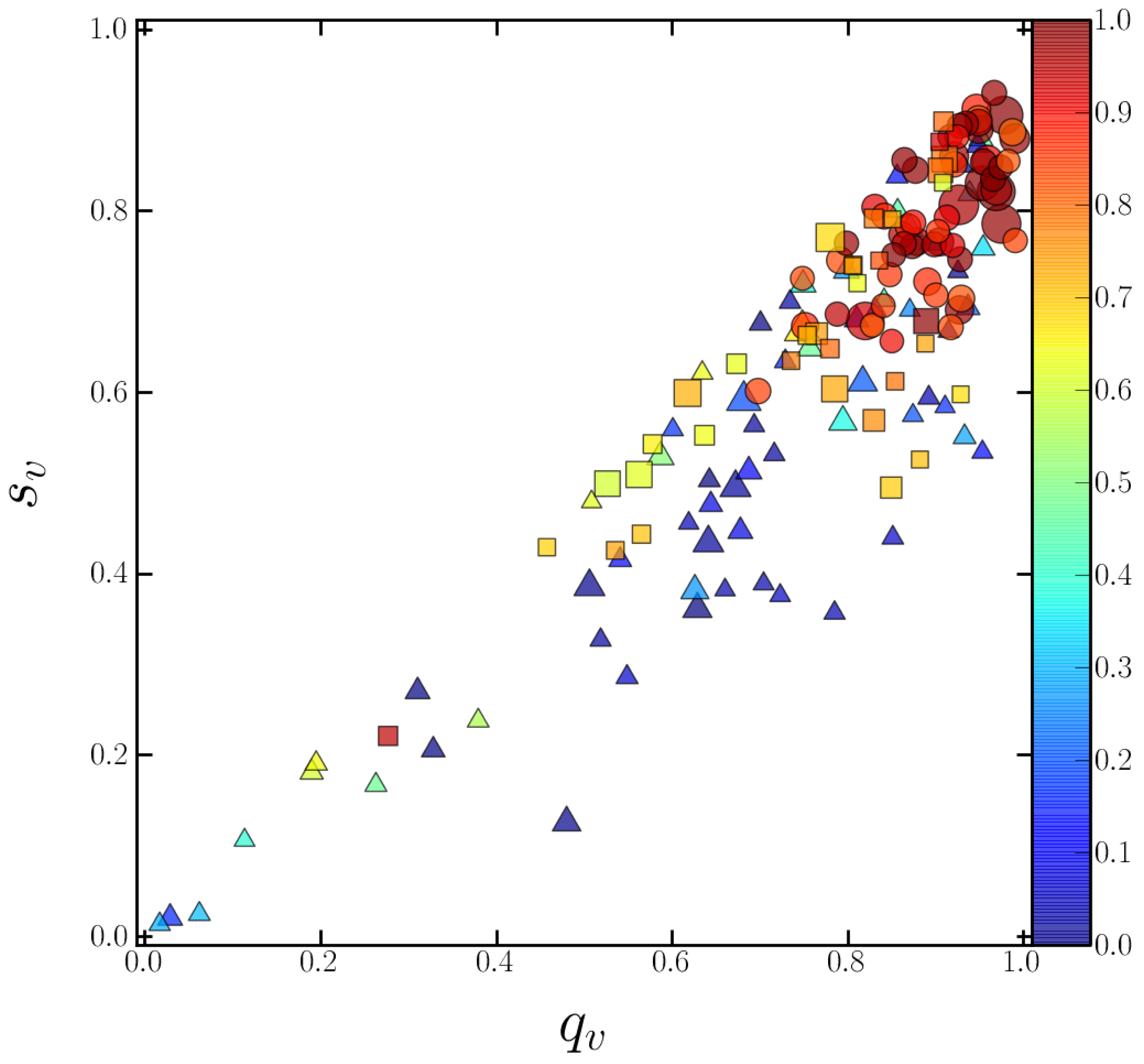}
    \caption{The dynamical state and phase-space morphology of substructures identified by \hst\ composed of $\geq500$ particles. We plot the axis ratios of the inertia tensor in the top panel and the velocity dispersion tensor in the bottom panel. Points are coloured according to $f_E$ and the marker size is proportional to $\log M_{\rm S}$, the logarithmic mass of the substructure. Marker styles indicate the type of substructure, with circles, squares, and triangles corresponding to subhaloes, transitional substructures, and tidal debris (streams, and shells) using the classification scheme outlined in \Eqref{eqn:morphclassification}. Note that $s\leq q$ \& $s_v\leq q_v$.}
    \label{fig:phasespacemorph}
\end{figure}
The substructure finders used in this study should, in principle, find all types of substructures from relatively compact, spherical, bound subhaloes to unbound shells and streams. Before we can examine how well substructure finders recovery tidal debris we first need to determine whether a substructure should be classified as tidal debris or a subhalo. We use the phase-space morphology, quantified using $q$, $s$, $q_v$, $s_v$, and the dynamical state, quantified by $\fracE$, discussed in \Secref{sec:pipeline}. The distribution of these key properties is shown in \Figref{fig:phasespacemorph} for a subset of substructures in the reference \hst\ catalogue. Recall that all the particle tracked by \hst\ originated from a distinct FOF halo, thus $\fracE$ is not affected by the inclusion of particles with high relative velocities, i.e.~false positives.

\par
This figure shows that there is a strong correlation between the inferred dynamical state and morphology and that substructures span the entire range of these quantities. Substructures with spherical phase-space morphologies, that is $q,s,q_v,s_v\sim1$, are generally self-bound with $\fracE\gtrsim0.8$, whereas highly anisotropic substructures typically have $\fracE\lesssim0.2$. There are a few substructures that appear to have spherical morphologies but are completely unbound. Visual inspection of these substructures show that these are shells or disrupted substructures that have filled quite a large region of their orbit space. We also see that most substructures have $s\sim q$ and that there are no substructures with $q\sim1$ and $s\lesssim0.5$. The absence of pancake like substructures indicates subhaloes are generally only elongated along one particular axis as they are tidally stretched.  
\begin{figure*}
    \centering
    \includegraphics[height=0.9\textheight]{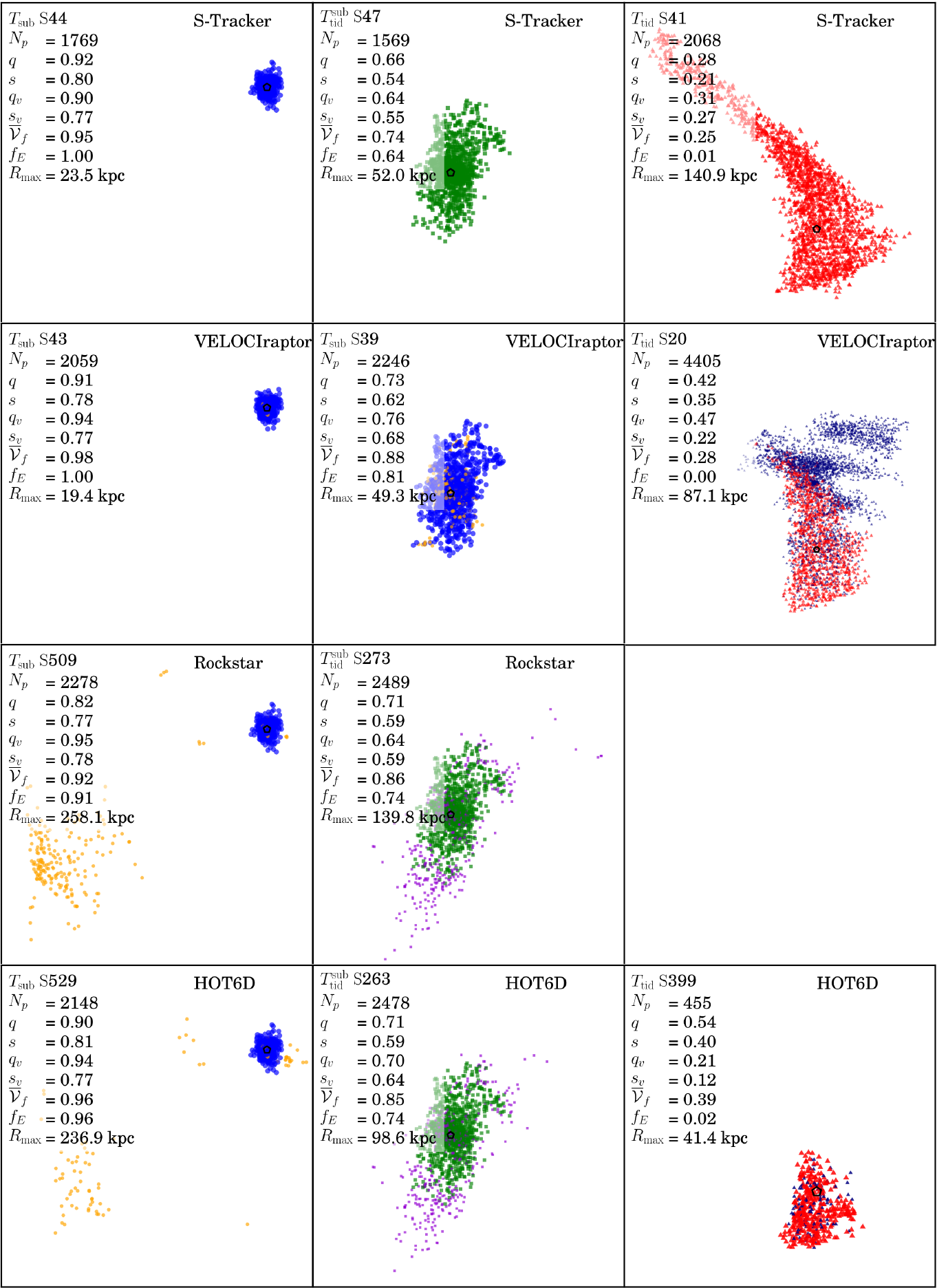}
    \caption{Example of substructures found by \hst\ (first row) and their counterparts identified by the non-tracking algorithms if a viable cross match is found (next three rows). We show three different classes of substructures, with a subhalo in the left panel, a transition substructure in the middle panel, and a tidally disrupted substructure in the right panel. We plot a projection of the particle distribution and show relevant structural quantities, plot an open pentagon for the centre-of-mass, along with the distance of the most distant particle to this point, $R_{\rm max}$. Note that particles are colour coded according to the morphological classification of the group, blue, green and red corresponding to $\Tsub$, $\Tsubtid$, and $\Ttid$ respectively. We also colour particles in the non-tracking codes that are not found in the \hst\ substructure to emphasize the false positives found, orange for subhaloes, purple for transition objects and navy blue for tidally disrupted objects. The scale is the same across finders (going down a column) but differs for each object (going across a row).}
    \label{fig:morphclass}
\end{figure*}

\par
For simplicity, we use three classes: subhaloes; tidal debris; and the transition region in between. We visually inspect a sample of 40 visually classified tidally disrupted substructures and 40 visually classified intact subhaloes chosen at random from our reference catalogue to determine these limits. We define intact subhaloes, transition substructures, and tidal debris as as $\Tsub$, $\Tsubtid$, and $\Ttid$ substructures respectively. The classification criteria, in part based on the work of \cite{elahi2011}, are listed below, 
\begin{subequations}
\label{eqn:morphclassification}
\begin{align}
    \Tsub:\,&
    \begin{array}{l}
      \left(\fracE>{\fracE}_{\rm sub}\right) \cap \\
      \left[(q\geq q_{\rm sub}) \cap (s\geq s_{\rm sub})\right] \cap \\ 
      \left[(q_v\geq q_{v,{\rm sub}})\cap(s_v\geq s_{v,{\rm sub}})\right]
    \end{array}\\
    \Tsubtid:\,&
    \begin{array}{l}
        (\neq \Tsub)\cap(\neq\Ttid)
    \end{array},\\
    \Ttid:\,&
    \begin{array}{l}
      \big\{
      \left({\fracE}_{\rm tid}<\fracE\leq{\fracE}^{\rm sub}_{\rm tid}\right) \cap \\
      \left[(q\leq q_{\rm tid}) \cup (s\leq s_{\rm tid})\right] \cup \\ 
      \left[(q_v\leq q_{v,{\rm tid}}) \cup (s_v\leq s_{v,{\rm tid}})\right]
      \big\}\cup\\ 
      \left(\fracE\leq{\fracE}_{\rm tid}\right)
    \end{array}
\end{align}
\end{subequations}
where the subscripts $_{\rm sub}$ \& $_{\rm tid}$ correspond to subhalo and tidal stream limits respectively. We find that
\begin{align}
    \begin{array}{lccccc}
      & \fracE & q& s & q_v& s_v\\
      {\rm sub:} & 0.80 & 0.70 & 0.60 & 0.60 & 0.40 \\
      {\rm sub-tid:} & 0.50 \\
      {\rm tid:} & 0.20 & 0.50 & 0.40 & 0.35 & 0.30
    \end{array}\notag
\end{align}
correctly classifies our test sample of substructures. This classification scheme is based primarily on the inferred dynamical state of the substructure, though we attempted to account the transition region using the phase-space morphology. These limits effectively state that if a substructure is generally a bound ellipsoid, its an intact subhalo. As the substructure becomes either more triaxial or less bound, the substructure should be considered a subhalo with tidal features. We consider even substructures that are {\em dominated} by an unbound component to be subhaloes with tidal features unless they have anisotropic phase-space morphologies, in which case they are considered tidal debris. Substructures that have very small bound mass fractions, $\leq20\%$, are considered tidal debris.

\par
An example of the substructures found and how they are classified is shown in \Figref{fig:morphclass}, along with the relevant parameters. These objects are chosen at random from the S-Tracker catalogue as stereotypical representatives of each class of substructure, with the tidal debris chosen to represent a tidal stream, i.e.~small $s$ and $s_v$ values. We also require objects to be composed of $>1000$ particles. The left panel shows a subhalo, the central panel a subhalo with tidal features, and the right panel shows a completely unbound tidal stream. We also show the best counterparts identified by \velociraptor, \rockstar, and \hot6d, and differentiate between particles belonging to the reference substructure from the \hst\ shown and other particles deemed by \hst\ to either belong to a different substructure or the background. 

\par
Before we discuss the statistical analysis of this comparison, its informative to visually examine the counterparts found by the non-tracking codes plotted in lower panels of \Figref{fig:morphclass}. The first obvious issue is that both of the phase-space finders, \rockstar\ \& \hot6d, have associated seemingly unrelated particles to S44, the subhalo, seen in the left column. These false positives occur because the separation between the two clusters of particles is effectively not significant with respect to the noise in the phase-space metric used, indicating the need for some post-processing. Subhalo candidates are pruned by removing ``unbound'' particles, which is not applicable here. 

\par
The subhalo with tidal features, S47 (middle column), is identified by all finders. However, this substructure contains several completely disrupted subsubstructures, which are not identified as separate substructures by any of the non-tracking codes (see discussion in \Secref{sec:trackingnotracking}). As a result, the fraction of bound particles is higher in the three non-tracking codes, such that \velociraptor\ classifies this object as a pure subhalo. \rockstar\ \& \hot6d\ also incorrectly associate outlying particles. 

\par
The picture is far more complicated for the very diffuse stream. It is only partially identified by \velociraptor \& \hot6d.  \velociraptor\ recovers a significant fraction of the stream, although the contamination appears high, having associated another substructure with this stream.  \hot6d only detects a small fraction. We also note that \Figref{fig:morphclass} indicates that a substructure's classification can change, i.e., $\Tsub\leftrightarrow\Tsubtid$ or $\Tsubtid\leftrightarrow\Ttid$, depending on the algorithm.

\subsection{General results}\label{sec:generalresults}
The total number of each type of substructure found by each finder is listed in \Tableref{tab:overallresults} along with the mass fraction in substructures. This table shows that the number of substructures identified varies greatly, by up to a factor of 1.6, though the total mass in substructures only differs by at most $20\%$. \hst\ finds the most distinct substructures, \hot6d finds the least. We discuss the reasons for these discrepancies below.
\begin{table*}
\centering\small
\caption{Basic properties of the substructures identified by each finder. Left column gives the finder name. The next columns show the total number of substructures found within the main halo composed of more than 20 particle $(N_{\rm S,tot})$ and the total fraction of mass in substructures for each type $(f_{\rm S,tot})$.}
\begin{tabular}{@{\extracolsep{\fill}}l| cccccccc}
\hline
\hline
    Finder Name & \multicolumn{2}{c}{All Substructures} & \multicolumn{2}{c}{Subhaloes $(\Tsub)$} & \multicolumn{2}{c}{Subhaloes+Tidal Features $(\Tsubtid)$} & \multicolumn{2}{c}{Streams \& Tidal Debris $(\Ttid)$} \\
     & $N_{\rm S,tot}$ & $f_{\rm S,tot}$ & $N_{{\rm S,tot}}$ & $f_{{\rm S,tot}}$ & $N_{{\rm S,tot}}$ & $f_{{\rm S,tot}}$ & $N_{\rm S,tot}$ & $f_{{\rm S},tot}$ \\
\hline
    \hst           & 3025 & 0.1776 & 483 & 0.0557 & 1628 & 0.0788 & 914 & 0.0422 \\
    \velociraptor  & 1953 & 0.1809 & 379 & 0.0949 & 1123  & 0.0599 & 388  & 0.0273 \\
    \rockstar      & 2330 & 0.2082 & 207 & 0.0655 & 1194  & 0.1110 & 929 & 0.0317 \\
    \hot6d         & 1883 & 0.1911 & 224 & 0.0700 & 846  & 0.0846 & 955 & 0.0365
\end{tabular}
\label{tab:overallresults}
\end{table*}

\par
We should emphasize that, despite the fact that \hst\ tracks particles, its catalogue is {\em not} necessarily complete. First, it relies on the standard FOF scheme to identify progenitor haloes. This scheme can artificially link haloes which are joined by bridges of particles. The two haloes in this FOF-structure will follow different orbits. Consequently, when \hst\ applies its phase-criterion it will find two groups, and as \hst\ only keeps the larger of the groups, only one will end up being tracked. Additionally, a subhalo can accrete mass from its host halo, especially while it still resides in the outskirts, though the amount of mass accreted is small \citep{hbt}. These newly accreted particles are never tracked as we do not allow substructures to be accreted from the background. Substructures can also merge, though this is rare \cite[][]{klimentowski2010}. If two substructures occupy the same phase-space volume for a dynamical time, should they be considered two separate substructures even if at some much early stage they originated from two different FOF haloes? There is not a simple answer to this question, hence for simplicity we do not merge substructures in \hst. Finally, due to the coarse grain search in the time domain for progenitor haloes, if a halo forms and is accreted between snapshots, these particles will never be tracked. For the data sets presented here, this only becomes an issue for haloes composed of $\lesssim50$ particles.

\subsection{How well is tidal debris recovered?}\label{sec:cross-matching}
\subsubsection{Cross-matching}
\begin{figure*}
    \centering
    \includegraphics[width=0.98\textwidth]{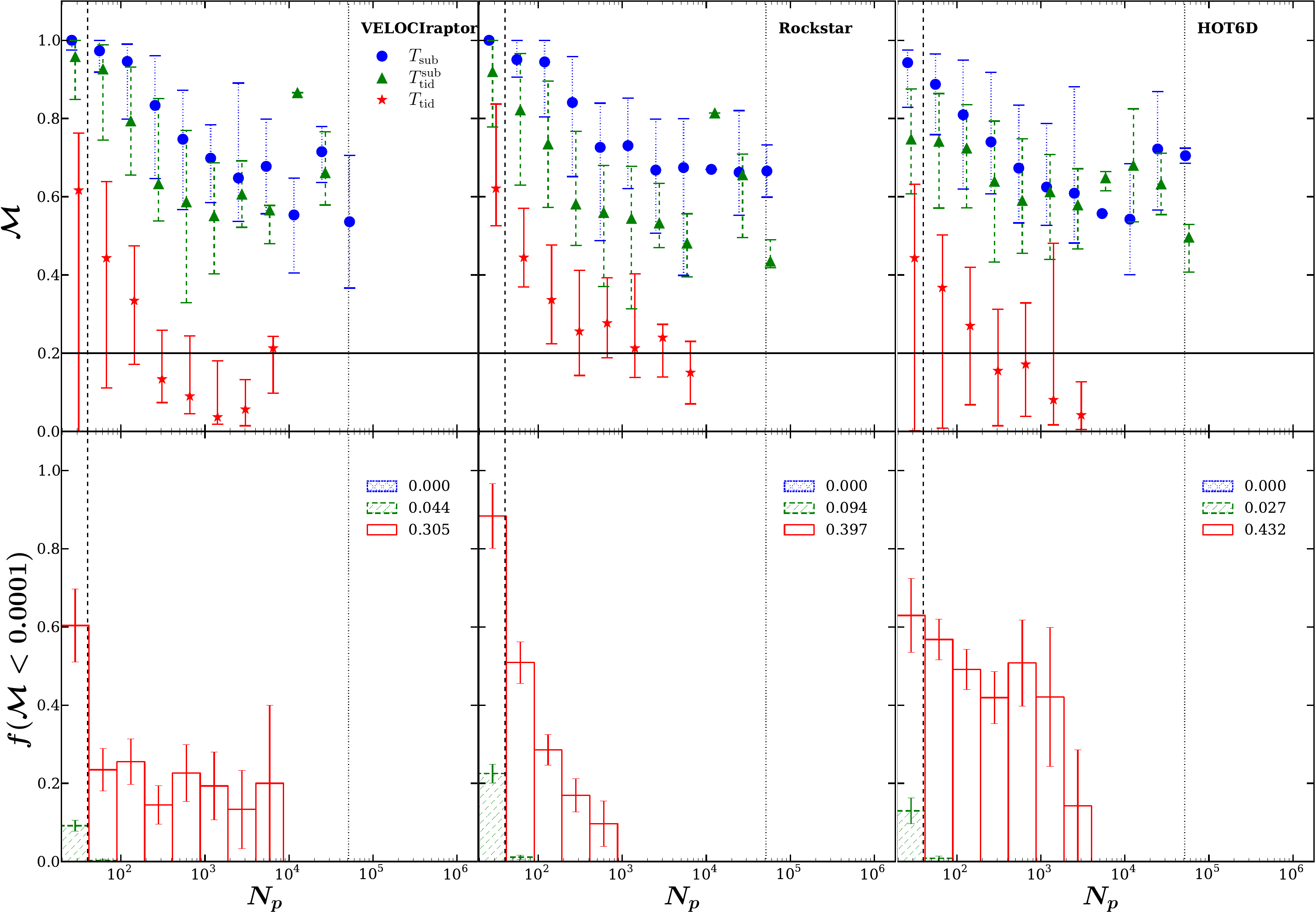}
    \caption{The top row shows the $20\%,80\%$ quantiles (error bars) and median (filled symbol) of the merit distribution of substructures in logarithmically spaced particle number bins categorized according to morphological type. Intact subhaloes are shown by (blue) circles and dotted lines, subhaloes with tidal features by (green) triangles and dashed lines, and tidal debris by (red) stars and solid lines. Note that for clarity we have offset these points by a small amount in each bin. The bottom row shows histograms of the fraction of all substructures {\em in each bin} that have $\merit<0.0001$, that is effectively no match is found, along with the Poisson error bars. Here subhaloes, transition substructures,  and tidal debris are shown by (blue) cross-hatched histogram, (green) hatched histogram, and a red open histogram respectively. In the legend we show the total fraction of substructures effectively not found in the reference catalogue, $f_{\rm tot}(\merit<0.0001)$. Here substructures are classified according to to the distribution of particles identified by the finder. Finally, we also plot a vertical line at 40 particles, twice the particle limit used to identify substructures, and the number particles at corresponding to $1\%$ of the halo's total, the point at which the largest substructure are often found.}
    \label{fig:merit}
\end{figure*}
  We quantify how well each finder does using three quantities, the merit, purity and recovery fraction. The merit of a cross match is 
\begin{equation}
  \mathcal{M}\equiv N^2_{\rm sh}/(N_{\rm 1} N_{\rm 2})
\end{equation}
where $N_{\rm sh}$ is the number of particles shared by the substructure in the comparison catalogue and with another substructure in the reference catalogue, and $N_{1}$ and $N_{2}$ are the number of particles in each substructure \citep{klimentowski2010,libeskind2010,knebe2013a}. As a substructure may share particles with several other substructures in the reference catalogue, we are only concerned with the maximum merit of a given substructure. We also calculate the total purity, $\mathcal{P}_{tot}$, of a substructure, which is the fraction of particles in a substructure that belongs to a substructure that originated from a distinct FOF-halo. Finally, we calculate fraction of a substructure in the reference catalogue recovered in {\em any substructure} in the comparison catalogue, $\mathcal{R}$, allowing a substructure to be split into multiple groups. Together, these quantify an individual match, the fraction of false positives, and the total fraction of a real substructure recovered. 

\begin{figure*}
    \centering
    \includegraphics[width=0.98\textwidth]{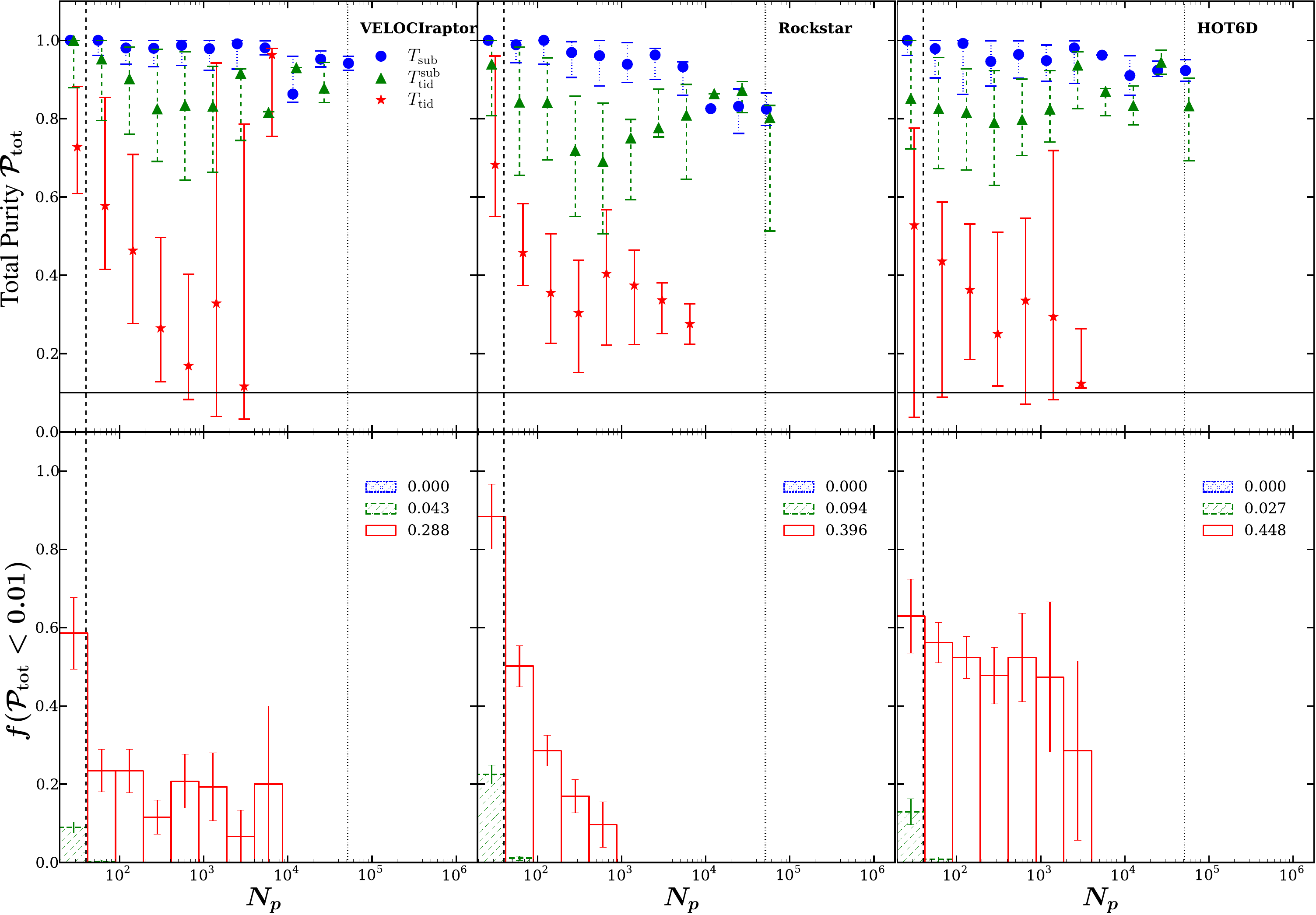}
    \caption{Similar to \Figref{fig:merit} but for the total purity distribution and the mass distribution of substructures that are effectively spurious, $f(\purity<0.01)$}
    \label{fig:purity}  
\end{figure*}
\begin{figure*}
    \centering
    \includegraphics[width=0.98\textwidth]{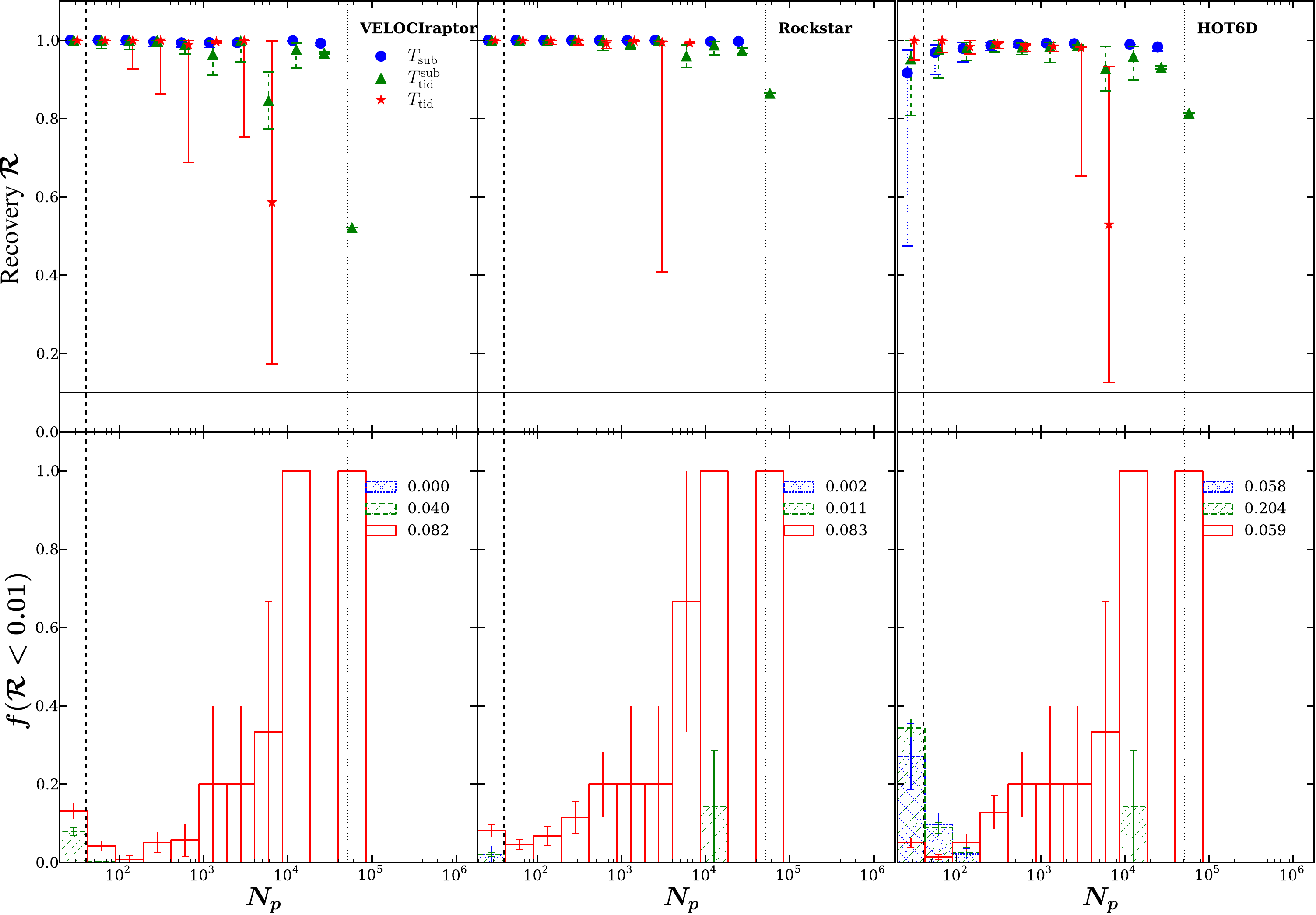}
    \caption{Similar to \Figref{fig:purity} but for the recovery fraction distribution. Here substructures are classified according to the substructure found by \hst, {\em not} the finders plotted.}
    \label{fig:recovery}
\end{figure*}

\par
The results are plotted in \Figref{fig:merit}-\ref{fig:recovery}, where we have categorized substructures according to the morphological types outlined in \Secref{sec:classification}. Note that substructures for \Figref{fig:merit} \& \ref{fig:purity} are classified according to the morphological type of the substructure found in the comparison catalogue, whereas in \Figref{fig:recovery} substructures are plotted according to their \hst\ morphological class. 

\par
In \Figref{fig:merit}, we show the median and quantiles of the merit distribution in several particle number bins for the three categories of substructures. \cite{libeskind2010} found that $\merit\sim0.2$ indicates a viable cross-match has been found. Here we deem any substructure with a merit of $\lesssim10^{-4}$ to be completely spurious. Substructures with $10^{-4}\lesssim\merit\lesssim0.2$ can be viewed as substructures that are poorly identified, though this simple interpretation is not always the case for tidal debris (see following discussion on \Figref{fig:tidaldebris}). If we compare the average merit value between substructure types, we clearly see that for all finders, less tidally disrupted substructures have higher $\merit$. More importantly, this figure indicates that subhaloes with tidal features, even those that are predominantly unbound are well matched by an object in the reference catalogue. Only a very small fraction of transitional substructures near the particle limit have no counterpart. 

\par
Of greatest relevance for this study is the fact that tidal streams and other diffuse unbound substructures are found with $\merit\gtrsim0.2$ for all three finders. It should be emphasized that due to the diffuse nature of streams and clouds, recovery can often be patchy, resulting in a stream being split into several different ``substructures'' resulting in a lower $\merit$. The greatest difference between the three comparison finders lie in the fraction of tidal streams not found in the reference catalogue. \velociraptor\ has the fewest number of substructures not well cross-matched to an individual substructure in the reference catalogue. Generally only streams consisting of $\sim20$ particles, near the number cutoff imposed, are not present in the reference catalogue, though there are some small fraction of spurious structures across a wide range of masses. \rockstar\ shows a strong resolution dependence on the number of spurious substructures, with a very large fraction occurring near the 20 particle limit imposed. This is not the case for \hot6d, where $\sim40\%$ of all very diffuse loosely bound substructures are spurious regardless of the number of particles the substructure is composed of. 

\par
Not all extra substructures in the non-tracking catalogues need be truly spurious detections. One possible source is from the fragmentation of a \hst\ substructure. This tracking code applies a phase-space window to link particles originating from a progenitor halo. If a fragment of a substructure at a given snapshot lies outside this window, it will no longer be tracked. However, given the large phase-space window used and the fact that most of these objects do not have a viable cross match in the other catalogues, suggest they are truly spurious substructures. Another potential source for these excess substructures are mergers. \cite{behroozi2012} found that particles can be excited to high kinetic energies and be ejected from the host halo during merger events. These particles do not constitute a coherent dynamical structure but due to their kinematic properties greatly differing from the background, they could be grouped together and identified by the non-tracking codes as a substructure. 

\par
The merit shows a mild inverse dependence with particle number, that is more massive substructures tend to have smaller merits. This trend is for $\Tsub$ \& $\Tsubtid$ substructures is primarily due to the presence of tidally disrupted subsubstructures and relates to when a substructure should be considered to have phase-mixed with its host. There is no simple answer to this question. In such instances, the non-tracking finders may identify a single substructure, whereas \hst\ may continue to track each substructure separately. Larger subhaloes are more likely to have accreted other substructures and better resolved subhaloes will have contained more substructure as a halo prior to be accreted themselves, hence the observed trend. The case is different for tidally disrupted substructures, where lower merits can occur when a single tidally disrupted substructure is split into several groups. Another complication is that tidally disrupted substructures have a higher likelihood of occupying the same space as another substructure due to their extended volumes. Consequently, non-tracking codes might consider the enveloped substructure as part of the tidal debris.

\par
The distribution of $\purity$, plotted in \Figref{fig:purity}, shows that subhaloes have very high purities. The mass dependence seen in \Figref{fig:merit} is no longer present, further indicating that part of the trend in $\merit$ is due to the grouping together or splitting of substructures. The small amount of contamination seen in subhaloes can partly be accounted for by the fact that \hst\ does not allow substructures to accrete particles from the background. Even the purity for subhaloes dominated by tidal features remains quite high at $\approx0.8$. Substructures classified as tidal debris a show a greater number of false positives, with purities around $\sim0.3-0.4$, though some tidal debris has purities as high as $0.8$. Note that the distribution of substructures with $\purity\lesssim0.01$ is similar to that of substructures with no viable individual cross-match seen in \Figref{fig:merit}.

\par
The recovery fraction plotted in \Figref{fig:recovery} shows similar trends to the two previous figures. Intact subhaloes and subhaloes with strong tidal features are almost completely recovered regardless of mass (save in the lowest mass bin where particle noise becomes an issue and for the largest subhalo which is dominated by very extended tidal features). Of the three finders tested, \hot6d\ has the strongest dependence on numerical resolution, even for subhaloes. The picture is not so clear when it comes to tidal debris, where several large $\Ttid$ substructures in the reference catalogue are missing in all the other catalogues. 

\subsubsection{``Missing'' substructures and disruption}\label{sec:missingobjects}
These findings initially suggest that, though the non-tracking codes are able to identify gravitationally unbound tidal debris with a reasonable amount of purity, the fraction recovered is surprisingly poor and {\em does not depend} on the size of the structure. However, a closer inspection reveals that these substructures are typically very diffuse clouds of particles that appear to have phase-mixed with the background. 

\par
\begin{figure}
    \centering
    \includegraphics[width=0.45\textwidth]{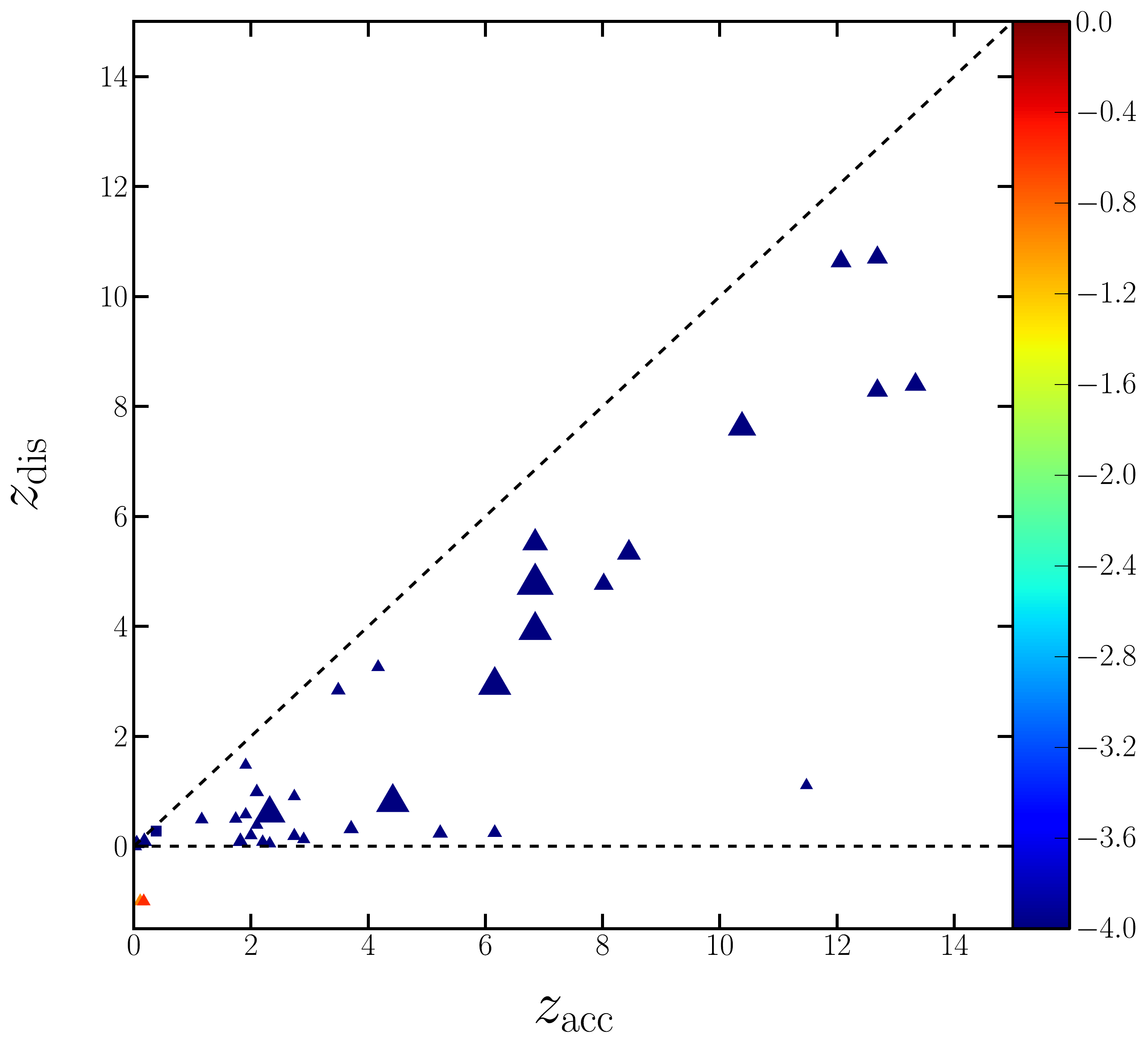}
    \caption{The accretion redshift and disruption redshift of substructures not recovered by \velociraptor. If a substructure is not considered disrupted we set its disruption redshift to -1. Circle, squares and triangles correspond to $\Tsub$, $\Tsubtid$ and $\Ttid$ substructures respectively. Points are colour coded according to $\log \left[M_{\rm b}(z=0)/M_{\rm FOF, infall}\right]$, the log of the fraction of self-bound mass at $z=0$ relative to the substructure's initial halo mass. Marker size scales with $\log M_{\rm S}$.}
    \label{fig:zaccretenorecovery}
\end{figure}

For instance, the largest tidally disrupted substructure in the reference catalogue that is not recovered by the other finders is composed of $\approx4\times10^4$ particles and was accreted at $z=6.85$. This substructure originated from an FOF halo was composed of $\approx 10^5$ and by $z=4.77$ less than $0.1\%$ of its original mass was self-bound. We plot the accretion and disruption redshift distribution of all missing substructures composed of more that 40 particles in \velociraptor\ catalogue in \Figref{fig:zaccretenorecovery}. The distribution for \rockstar\ and \hot6d for the large missing tidal debris substructures are similar. Here we define the disruption redshift as the redshift at which less than $0.001$ of a substructures original mass is self-bound. We see that the accretion and disruption redshifts for these substructures is $z_{\rm acc}\gtrsim4$ and $z_{\rm dis}\gtrsim2$ respectively. Only a few substructures composed of $\sim40$ particles are not found and also not completely disrupted by $z=0$. Most of these missing substructures typically have self-bound masses of $\lesssim10^{-4}$ times their initial infall FOF mass. We argue that these substructures should no longer be tracked.

\subsubsection{Tracking vs Non-tracking codes} \label{sec:trackingnotracking}
Let us now examine the details of the cross-matched examples in \Figref{fig:morphclass} listed in \Tableref{tab:groupexample}. The cross-matched substructures for the subhalo and transition substructure in the reference catalogue, S44 \& S47, all contain matches to other substructures in the reference catalogue. These ``extra'' substructures are in fact unbound clouds of particles that reside in the substructure, that is tidal debris of a subsubhalo residing in a subhalo, which appear to have phase-mixed with the background subhalo. Consequently, these substructures are not identified as distinct substructures, though they are completely recovered. A large fraction of the $\Ttid$ with large recovery fractions fall into this category. This is the reason that the merit seen in \Figref{fig:merit} has a dependence on particle number. We note that for S44, only three smaller substructures in the \hst\ catalogue appear to reside in the phase-space volume occupied by S44. The extra substructures associated by the two phase-space based codes, \rockstar\ \& \hot6d, reside in the patch of unassociated particles seen in \Figref{fig:morphclass}. Only small fractions of these substructures are recovered, hence the low $\overline{\mathcal{R}}_{\rm e}$. 
\begin{table}
\centering\small
\caption{Statistics of cross-matched groups in \Figref{fig:morphclass} relative to the \hst\ groups. We list the merit, total purity, recovery fraction, the extra number of reference substructures in the cross matched substructure ($n_{\rm e}$), and the average recovery fraction of these extra matches, ($\overline{\mathcal{R}}_{\rm e}$).}
\begin{tabular}{@{\extracolsep{\fill}}cl ccc ccc}
\hline
\hline
    Reference & Finder & $\merit$ & $\purity$ & $\recoveryfrac$ & $n_{\rm e}$ & $\overline{\mathcal{R}}_{\rm e}$ \\
    Substructure& \\
\hline
    \multirow{3}{*}{$\Tsub$ S44} 
    & \velociraptor & 0.84 & 1.00 & 0.99 & 3 & 1.00 \\
    & \rockstar     & 0.78 & 0.92 & 1.00 & 4 & 0.78 \\
    & \hot6d        & 0.82 & 0.97 & 0.99 & 5 & 0.65 \\
\hline
    \multirow{3}{*}{$\Tsubtid$ S47} 
    & \velociraptor & 0.65 & 0.97 & 0.96 & 7 & 1.00 \\
    & \rockstar     & 0.61 & 0.89 & 0.99 & 7 & 0.99 \\
    & \hot6d        & 0.61 & 0.89 & 0.98 & 7 & 0.99 \\
\hline
    \multirow{3}{*}{$\Ttid$ S41} 
    & \velociraptor & 0.16 & 0.26 & 0.60 & 2 & 1.00 \\
    & \rockstar     & NA \\
    & \hot6d        & 0.13 & 0.76 & 0.17 & 0 & NA \\
\end{tabular}
\label{tab:groupexample}
\end{table}

\par
These ``extra'' substructures raise a question for codes which track tidal debris: at what point should they be considered part of the background? Substructures that are indistinguishable from the {\em halo} background are typically disrupted at high $z$ (see \Figref{fig:zaccretenorecovery}). This is not always the case for these structures, though they typically have $\fracE\lesssim0.2$. However, given that these substructures occupy similar phase-space volumes to their host {\em (sub)halo} and are loosely bound, it is an open question whether they should be considered separate substructures at all. The phase-mixed substructures account for a significant fraction of the differences in the number of substructures identified by \hst's and the non-tracking codes, if one accounts for the spurious $\Ttid$ substructures in non-tracking catalogues (see \Tableref{tab:overallresults}).


\par
Finally, we turn our attention to the stream S41 in \Figref{fig:morphclass}, which appears to be the prototypical tidal stream in the reference catalogue. The \velociraptor\ \& \hot6d\ counterparts to the diffuse stream, S41, highlight the difficulty identifying streams. Both finders associate small fractions ($\sim1\%$) of the particles \hst\ associates with this stream with other substructures. \velociraptor\ and \hot6d\ contaminant one substructure with these particles where these particles comprise a significant fraction of the substructure ($\sim10\%$). Only a small fraction is recovered by \hot6d counterpart, but the purity is high, whereas \velociraptor\ recovers a significant fraction but is contaminated by background halo particles and particles belonging to another completely disrupted substructures that intersects this stream. \rockstar\ does not find a viable counterpart, possibly because the structure is very diffuse and elongated in configuration space. However, it does recover a very small fraction ($\sim0.03$) of this tidal debris but associates these particles with a $\Tsubtid$ substructure, which is an excellent cross-match to a different substructure in the \hst\ catalogue. 

\par
\begin{figure*}
    \centering
    \includegraphics[width=0.98\textwidth]{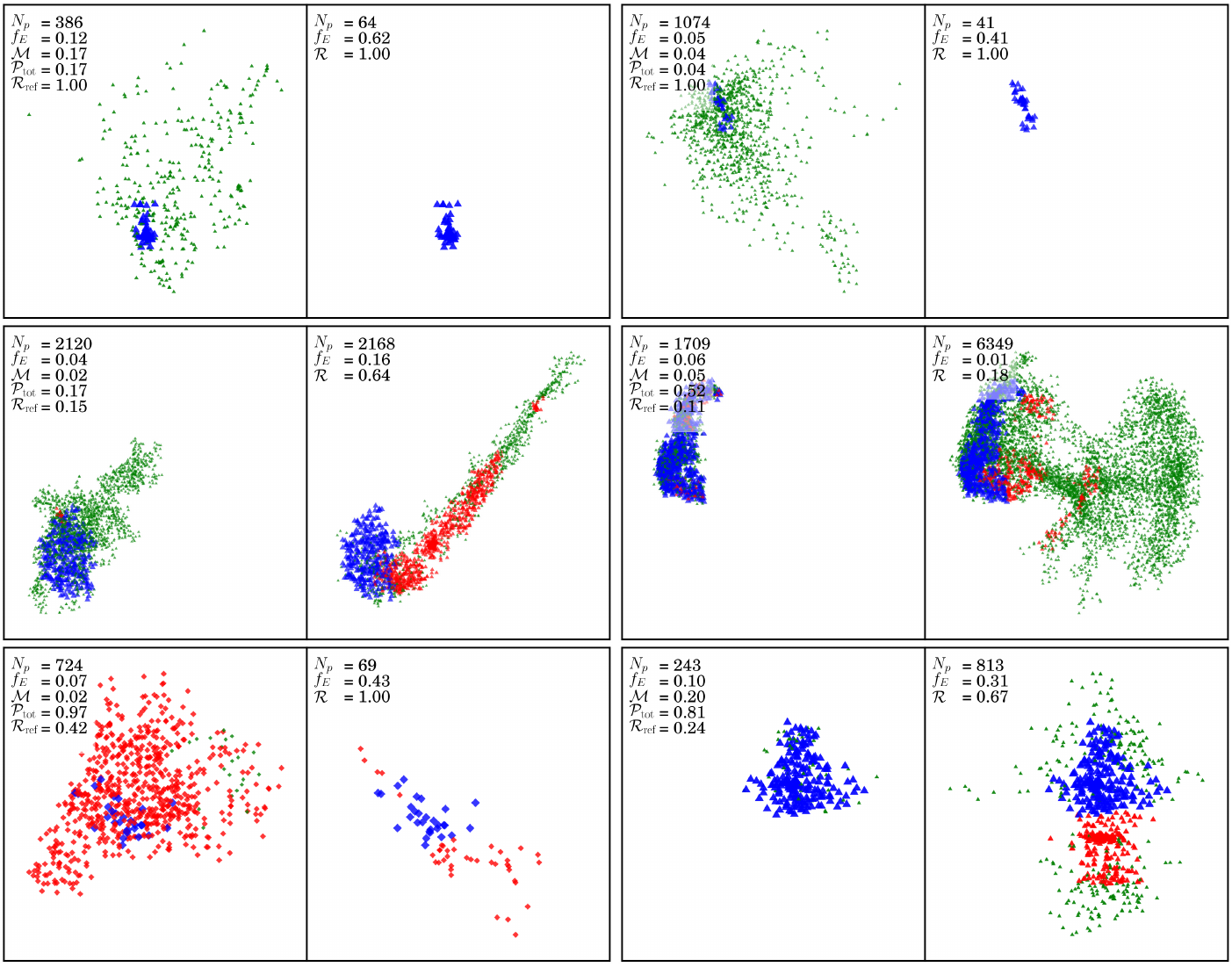}
    \caption{Examples of tidal debris found by non-tracking codes (left-side panel) an the corresponding reference substructure with the highest merit in the \hst\ catalogue (matching right side panel). Particles in both the tidal debris example and the reference substructure are plotted in blue, particles belonging to a substructure other than that with the highest merit are plotted in red, and particles associated with the background by either the non-tracking code or \hst\ are plotted in green. For the substructure found by the non-tracking code, we show the number of particles in the substructure $N_p$, $\fracE$, $\merit$, $\purity$, and the fraction of the reference substructure recovered $\mathcal{R}_{\rm ref}$. For the reference substructure we show $N_p$, $\fracE$, and $\recoveryfrac$, the total fraction recovered across all substructures in the non-tracking catalogue.}
    \label{fig:tidaldebris}
\end{figure*}

The tidal debris example in \Figref{fig:morphclass} and the trends seen in \Figref{fig:merit}-\Figref{fig:recovery} indicate that identifying {\em almost completely disrupted} substructures poses a special challenge. There appears to be tidal debris with low merits but high purities or low merits but high recovery fractions. To explore these issues we show a few tidal debris examples with a variety of $\merit$, $\purity$ and $\recoveryfrac$ values in \Figref{fig:tidaldebris}. We emphasize that this figure is not a representative sample of the population of tidally disrupted substructures found by the non-tracking codes. It merely highlights different scenarios.

\par
The first row shows two examples where the non-tracking algorithms perform poorly, associating background particles in the region surrounding a small substructure, which is {\em becoming tidally disrupted}. These incorrectly associated particles cause the substructure to be considered $\Ttid$ instead of $\Tsubtid$. These examples have low $\merit$, $\purity$ but the entire substructure is recovered.

\par
The examples in the second row are of substructures which recover a small fraction of a tidally disrupted substructure in the \hst\ catalogue. In the example in the left column, both $\merit$ \& $\purity$ are low but this is not a spurious substructure. Most of the tidally disrupted reference substructure is recovered but has been split into several groups, each with low $\merit$. In the example in the right column, the substructure has a low merit as only a small fraction of the large tidally disrupted reference substructure has been recovered by this substructure. However, the substructure in the non-tracking catalogue is not severely contaminated by the background. 

\par
The last row shows two different cases. In the left column, the substructure is very pure but has a very low merit. In fact, the reference object with the highest merit is {\em not} the object that should be associated with this reference substructure. Most of the particles here originate from the tidal tails of a much larger subhalo composed of $\approx39000$ particles. The tidal tails of this large reference substructure happen to intersect this smaller substructure, which is in the process of being tidally disrupted. The right column shows tidal debris object with high merit (recall $\merit\gtrsim0.2$ indicates a good counterpart has been identified),  high purity but low recovery fraction. Here the reference substructure has been split in two, a trailing stream and a leading stream which contains the progenitor subhalo.

\subsection{Summary of Comparison} \label{sec:comparisonsummary}
In summary, all three non-tracking finders identify tidal debris.  The statistics of each finder is listed in \Tableref{tab:comparisonsummary}, where the average $\merit$, $\purity$, and $\recoveryfrac$ listed are calculated {\em excluding} the contribution from spurious objects and only for substructures composed of more than 40 particles in order to minimize the effects of poorly resolved substructures. First, we should emphasize that all finders perform remarkable well in identifying subhaloes with significant amounts of mass in unbound tidal features. However, the performance of these finders degrades as substructures become increasingly unbound. Focusing on the statistics of tidally disrupted substructures, we see all the non-tracking finders on average identify viable cross-matches, i.e.~$\merit\gtrsim0.2$, though they all also miss a small fraction of the tidally disrupted substructures present. \rockstar\ has the highest recovery fraction and the highest average merit. \velociraptor\ has the highest overall purity, the fewest completely missing objects and the fewest spurious objects. \hot6d\ has the largest number of spurious unbound substructures and also suffers from numerical resolution issues at particle numbers of $\lesssim100$, whereas the other two suffer from strong resolution effects for substructures composed of $\lesssim50$ particles. All non-tracking codes unfortunately identify a non-negligible fraction of spurious objects. These spurious objects and the significantly lower purity suggest that some post-processing may be required, similar in purpose to the unbinding routines used for subhaloes. It is possible that his number can be reduced at the cost of decreasing $\recoveryfrac$ by using more conservative selection criteria \cite[see discussion in][]{elahi2011}. 

\begin{table}
\centering\small
\caption{Overall performance of the non-tracking finders based on substructures composed of $\geq40$ particles (twice the particle limit used to find substructures). We show the average merit, purity, recovery fraction {\em excluding} spurious objects, the fraction of spurious detections, $f_{\rm spur}$, and the fraction of missing substructures, $f_{\rm miss}$. The fraction of missing substructures is corrected for the substructures that have phase-mixed with the background by discarding any substructure in the \hst\ catalogue which has $z_{\rm dis}>0.2$, decreasing this fraction by $0.03$.}
\begin{tabular}{@{\extracolsep{\fill}}ll ccccc}
\hline
\hline
    Finder & Type & $\overline{\merit}$ & $\overline{\purity}$ & $\overline{\recoveryfrac}$ & $f_{\rm spur}$ & $f_{\rm miss}$ \\
\hline
    & $\Tsub$ &     0.81 & 0.96 & 0.99 & 0.00 & 0.00 \\
    \velociraptor
    & $\Tsubtid$ &  0.79 & 0.87 & 0.98 & 0.00 & 0.002 \\
    & $\Ttid$ &     0.25 & 0.43 & 0.93 & 0.21 & 0.01 \\ 
\hline
    & $\Tsub$ &     0.80 & 0.94 & 0.99 & 0.00 & 0.00 \\
    \rockstar
    & $\Tsubtid$ &  0.72 & 0.79 & 0.99 & 0.00 & 0.001 \\
    & $\Ttid$ &     0.36 & 0.36 & 0.96 & 0.25 & 0.03 \\ 
\hline
    & $\Tsub$ &     0.71 & 0.95 & 0.95 & 0.00 & 0.04 \\
    \hot6d 
    & $\Tsubtid$ &  0.68 & 0.80 & 0.93 & 0.003 & 0.07 \\
    & $\Ttid$ &     0.26 & 0.36 & 0.95 & 0.38 & 0.03 \\ 
\end{tabular}
\label{tab:comparisonsummary}
\end{table}

\par
We note that due to the particle number trends seen in \Figref{fig:merit}-\ref{fig:recovery}, particularly for $\merit$ \& $\purity$, which decrease slightly with increasing particle number in large tidal debris, the exact values listed in the table should be treated with some caution. The increasing volume occupied by tidally disrupted substructures means that lower merits and purities are not unexpected. These structures are more likely to be split into several groups and may envelop other tidal debris, as illustrated in \Figref{fig:tidaldebris}. Nevertheless, these values are useful for gauging the performance of each code, the incompleteness of the catalogues and the contamination present.

\section{The Substructure Distribution}\label{sec:substructureprop}
\subsection{Mass function}\label{sec:massfunc}
\begin{figure}
    \centering
    \includegraphics[width=0.45\textwidth]{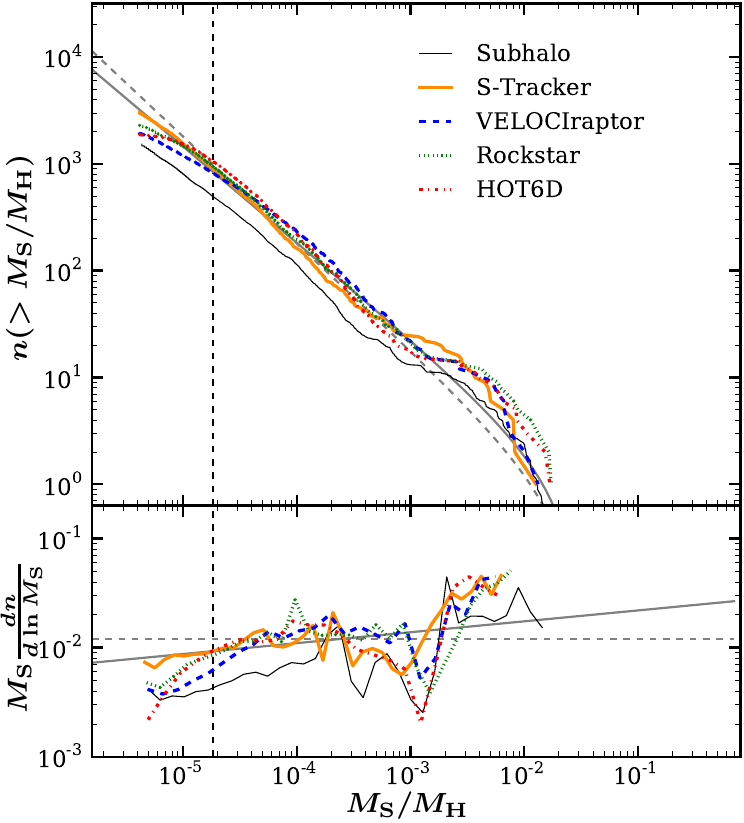}
    \caption{The cumulative substructure mass function in shown in the upper panel and the differential mass function in the lower panel. The dashed vertical line shows the mass-scale of substructures composed of 100 particles. Note that the differential mass function in the lower panel has been weighted by $M_{\rm S}$ to compress the vertical scale and highlight the power-law index of the mass function. For comparison, we show the average {\em subhalo} mass function from \protect\cite{onions2012} and the prediction from \Eqref{eqn:massfunc} for $\alpha=0.9$ \& $\alpha=1.0$, shown by the solid gray and dashed gray lines respectively.}
    \label{fig:massdistrib}
\end{figure}

The simplest distribution to examine is the general substructure mass distribution, shown in \Figref{fig:massdistrib}. A subhalo's mass is {\em not} very well defined and the dynamical mass of a stream is even less well defined, especially since the recovery can be patchy. However, at least for \hst's catalogue, this distribution is effectively the mass distribution of progenitor haloes modified partially by a tidal field. Before we discuss this figure, it is important to recall some salient facts about the {\em subhalo} mass function. The differential subhalo mass function is well characterised by power-law, e.g.
\begin{align}
  \frac{dn}{d\ln M_{\rm sub}}=AM_{\rm sub}^{-\alpha}\label{eqn:massfunc}
\end{align}
where $\alpha\approx0.9$. \cite{onions2012} \& \cite{knebe2013a} found that different finders recover the same cumulative subhalo mass function with a scatter of $20-30\%$ and individual substructures with a scatter of $\sim3\%$. 
\par
This figure shows that, despite differences in how substructures are identified, these algorithms give similar mass functions. Nevertheless, there are some key differences. \hst\ identifies significantly more substructures but most of these structures are composed of $\leq100$ particles. A careful visual inspection of these excess substructures indicates that most are best described as unbound collections of shells or clouds of particles, often residing in another subhalo as discussed previously. Both \rockstar\ \& \hot6d\ find significantly more substructures than \velociraptor\ at low particle numbers, but many of these structures appear to be numerical artefacts (see \Figref{fig:merit}). 

\par
There are also a few differences at the high mass end. Here subhaloes containing completely disrupted subsubstructures have larger masses in the non-tracking codes since these codes cannot separate the subsubstructure from the background substructure and we only allow particles to belong to one substructure (see discussion in \Secref{sec:cross-matching}). There are also several tidal debris substructures in the \hst\ catalogue not recovered by the other finders as they have effectively phase-mixed with the background.

\par
Finally, numerous studies show that the differential {\em subhalo} mass function has a power-law index that is $\alpha<1$ but what about the {\em substructure} mass function? A power-law of one is a special case where the mass in substructures is logarithmically divergent when extrapolated to arbitrarily small masses. The divergence cannot occur physically as a sharp cut-off in the halo mass spectrum is expected at the thermal dark matter free-streaming scale. Such an index would imply that there is in fact no smooth halo and all of the mass is contained in dynamically distinct substructures down to the free stream scale of dark matter, where structures are composed of the fundamental dark matter streams \citep{vogelsberger2010,abel2012}. 

\par
In the lower panel of \Figref{fig:massdistrib}, we plot the mass weighted differential specially to explore this issue. The data is too noisy to make conclusive statements but we do see that the \hst\ differential {\em substructure} mass function appears to be flatter than the differential {\em subhalo} mass function. Unfortunately, given the resolution of A-4, we cannot comment on the degree of change. These results, along with those of \cite{maciejewski2011}, indicate that mergers and tidal disruption modify the substructure distribution in such a fashion as to flatten the mass function.

\subsection{Velocity dispersion and distinguishing between streams and clouds} \label{sec:veldispfunc}
\begin{figure}
    \centering
    \includegraphics[width=0.45\textwidth]{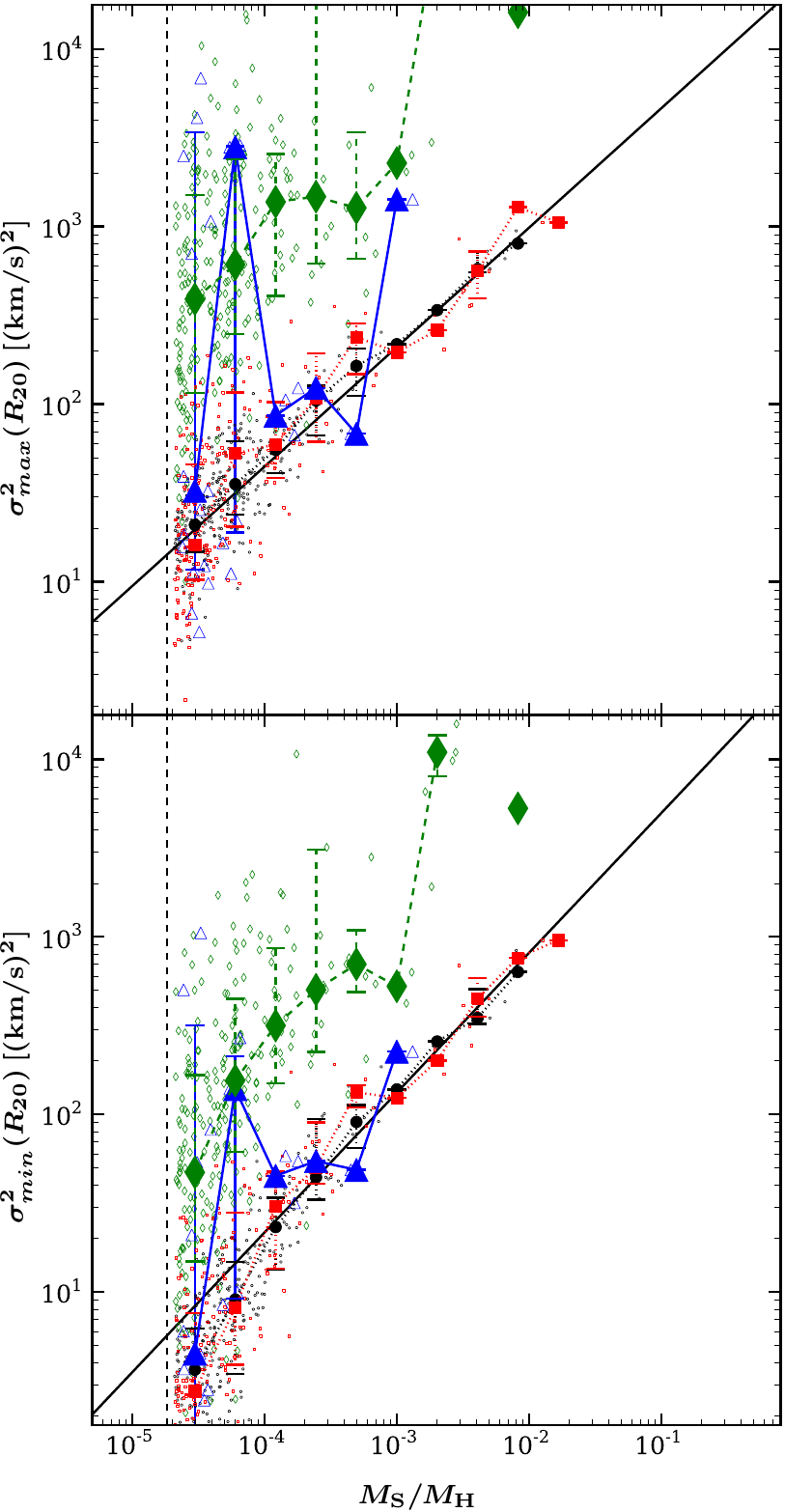}
    \caption{The relation between velocity dispersions and mass. We show the quantiles of the distribution in logarithmically spaced mass bins categorized according to morphological type (filled points and error bars denote the same quantiles shown in \Figref{fig:merit}). Here intact subhaloes are shown by (black) circles and dotted lines, subhaloes with tidal features by (red) squares and dotted lines, tidal streams by (blue) stars and solid lines, and clouds by (green) diamonds and dashed lines. We also plot the individual data points using small open markers. Finally, we also plot a fit to the $\Tsub$ \& $\Tsubtid$ data (solid black line).}
    \label{fig:massveldispersion}
\end{figure}
The ambiguity of defining an edge for (sub)haloes means $M_{\rm S}$ is not the ideal quantity for characterizing the the subhalo distribution. The situation is even worse when one includes tidal debris. The recovery of tidal debris is patchy. An excellent example is shown in in the bottom right panel of \Figref{fig:tidaldebris}, where the subhalo undergoing tidal disruption in the \hst\ catalogue is split into two groups by a non-tracking code. For (sub)haloes, a more physically meaningful quantity commonly used is the maximum circular velocity, $V^2_{\rm max}=GM/R_{\rm max}$, which is observable and contains dynamical information. $V_{\rm max}$ is also less prone to scatter ($\lesssim1\%$ \& $\sim10\%$ for individual substructures and the full cumulative distribution respectively, see \citealp{knebe2013a}). However, for completely disrupted, aspherical substructures, it is not a physically meaningful quantity. The velocity dispersion fortunately meets these criteria. 

\par
Structures in virial equilibrium have velocity dispersions which depend on the enclosed dynamical mass, i.e. $\sigma^2\propto M/r$. Specifically, one can show for a spherical system using the Jeans equation that the {\em local} radial velocity dispersion at a radius $r$ is
\begin{align}
  \sigma^2_{r}(r)=\frac{3V_c(r)^2}{5\frac{d\ln\rho}{d\ln r}-2\frac{d\ln Q}{d\ln r}+6\beta(r)},
\end{align}
where $V_c(r)^2=GM(r)/r$ is the circular velocity, $Q\equiv\rho/\sigma^3$ is the pseudo-phase-space density \cite[e.g][]{taylor2001,ludlow2010} and $\beta\equiv1-(\sigma^2_\theta+\sigma^2_\phi)/2\sigma^2_r$ is the anisotropy parameter. Assuming the physical and pseudo-phase-space density profiles are power laws, $\gamma\equiv -d\ln\rho/d\ln r$ and $\alpha\equiv -d\ln Q/d\ln r$, with minimal dependence on $r$, one can show that the average velocity dispersion in a sphere of radius $R$ is 
\begin{align}
  \sigma^2_r(R)&=\frac{\int\rho(r)\sigma_r^2(r) r^2dr}{\int\rho(r)r^2dr},\notag\\
  &=\frac{3GM(R)}{2R(5-2\gamma)}\left[\frac{1}{2\alpha+6\beta-5\gamma}\right],\notag\\
  &=\frac{3V_c^2(R)}{(5-2\gamma)}\left[\frac{1}{2\alpha+6\beta-5\gamma}\right].
\end{align}
For spherical (sub)haloes with NFW density profiles, this is a reasonable approximation as $\alpha\approx1.875$ and $\gamma$ slowly increases monotonically from 1 in the inner regions to 3 in the outer regions. In this case this dispersion peaks at the same radius as the rotation curve, $\approx2r_s$, the NFW scale radius. This radius, $R_{20}$, also encloses $\sim20\%$ of the mass, assuming the (sub)halo is truncated at $M_{200}$.

\par
Unfortunately, we are not dealing with spherically symmetric subhaloes but triaxial subhaloes and physically extended tidal debris. We therefore calculate the velocity dispersion tensor using the inner $20\%$ of the particles, the same region that is used to calculate the centre-of-mass, using a bi-weight estimator. By using a small volume centred on the densest and most bound part of a substructure, we minimize the contribution of orbital gradients and the tidally heated regions of the substructure. The bi-weight estimator also reduces the estimator's sensitivity to outliers. The dispersion tensor accounts for the triaxiality. Here we focus on the maximum \& minimum dispersions $\sigma^2_{max}$ \& $\sigma^2_{min}$ of the tensor within $R_{20}$.

\par
We explore the correlation of the {\em central} velocity dispersion with the dynamical mass of subhaloes, whether they are intact or they are dominated by tidal features in \Figref{fig:massveldispersion}. Note that here we plot these quantities for the \hst\ catalogue only. We see that $\sigma^2_{max}$ \& $\sigma^2_{min}$ of $\Tsub$ {\em and} $\Tsubtid$ substructures are strongly correlated with the current dynamical mass, with Pearson coefficient of $\approx0.94$. A fit to the data, $\log \sigma^2 = A+\alpha\log (M_{\rm S}/M_{\rm H})$, yields $\alpha=0.66\pm0.09$ and $0.86\pm0.09$ for the largest and smallest velocity dispersions respectively. Even for substructures with $\fracE\approx0.5$, that is {\em predominantly unbound}, the velocity dispersion is a useful proxy for the total dynamical mass. The scatter here arise in part due to the inherent scatter in triaxiality of velocity ellipsoid and the radius at which the rotation curve peaks relative to the substructures virial radius, which are related to the observed scatter in the density concentration parameter and the triaxiality of haloes \cite[see for instance][]{maccio2007,neto2007}.

\par
The picture is more complicated for tidal debris. As a subhalo orbits its host, particles outside the tidal radius become unbound. The tidal radius of a subhalo with mass $M_{\rm S}$ at a distance $R$ from the host's centre-of-mass with a velocity $V$ is $r_{\rm tid}\sim R\left[M_{\rm S}/M_{\rm H}(R)\right]^{1/3}$. Unbound particles originating from a progenitor subhalo will be spread in configuration and velocity-space on scales of $r_{\rm tid}$ and the dispersion at that radius, where the fractional spread of these particles is approximately equal in both space and velocity (see \citealp{binneytremaine2008,helmi1999,tremaine1999} for discussion in spread in angle-action space, and \citealp{knebe2005} for a discussion in the spread of integrals-of-motion space inferred from cosmological simulations). For elliptical orbits in a constant spherical potential, these particles will spread along the orbital plane of the subhalo while the width and dispersion orthogonal to the orbital plane remain roughly constant \cite[e.g.][]{helmi1999,johnston2001}. The evolution of the physical and velocity dispersions parallel to the orbital plane within a {\em region} is not as clear as the measured dispersion will contain contributions from orbital gradients.

\par
This picture is complicated by the fact that in our cosmological simulations, the potential is not constant, nor spherical and subhaloes have a variety of orbits, resulting in many forms of tidal debris, from shells or clouds (see for instance, \citealp{johnston2008}, who studied the properties in non-cosmological simulations of spherical subhaloes falling into a time-dependent potential containing a bulge, disk and spherical halo, and \citealp{warnick2008}, who examine tidal debris in fully cosmological simulations). Amorphous clouds of tidal debris will have high inferred velocity dispersions that are no longer correlated to their mass as these structures are phase-mixing with the background. However, stream-like substructures should have velocity dispersions orthogonal to their orbital plane that have not been significantly heated and lie near the line defined by subhaloes. 

\par
To examine whether streams follow the velocity dispersion relation of subhaloes but shells/clouds do not, we sub-classify tidal debris into stream-like and cloud-like morphologies. A substructure is considered to have a cloudy morphology if $s\geq s_{\rm sub}$ or $s_{v}\geq s_{v,{\rm sub}}$. Using this classification scheme, the \hst\ tidal debris substructures are split into 247 $(27\%)$ stream-like substructures and 667 $(73\%)$ cloud-like substructures. 

\par
The dispersions of this tidal debris are also plotted in \Figref{fig:massveldispersion}. This figure also shows that the dynamical state greatly affects this correlation. Tidal debris that does not appear stream like, i.e.~well confined to an orbital plane with small $s$ \& $s_v$ values, are significantly hotter, exhibit greater scatter and do not follow the subhalo correlation. These high dispersions indicate that these structures have evolved under the influence of the host halo's tidal field for several dynamical times. Their amorphous morphologies and high dispersions is why it is difficult to recovery these substructures without using temporal information (see discussion in \Secref{sec:missingobjects}).

\par 
However, stream-like substructures do follow the minimum dispersion correlation observed for subhaloes, indicating that this quantity is still a useful proxy for mass. The larger velocity dispersion shows significantly more scatter due to the stretching of the substructure along the orbital plane. There are several outliers that are classified using this scheme as stream-like substructures with dispersions that are a factor of $10-100$ times higher than the subhalo prediction, indicating either that our classification scheme is not perfect or that these stream-like substructures have been in heated in some way, such as by a close encounter with another substructure. 

\par
\begin{figure}
    \centering
    \includegraphics[width=0.45\textwidth]{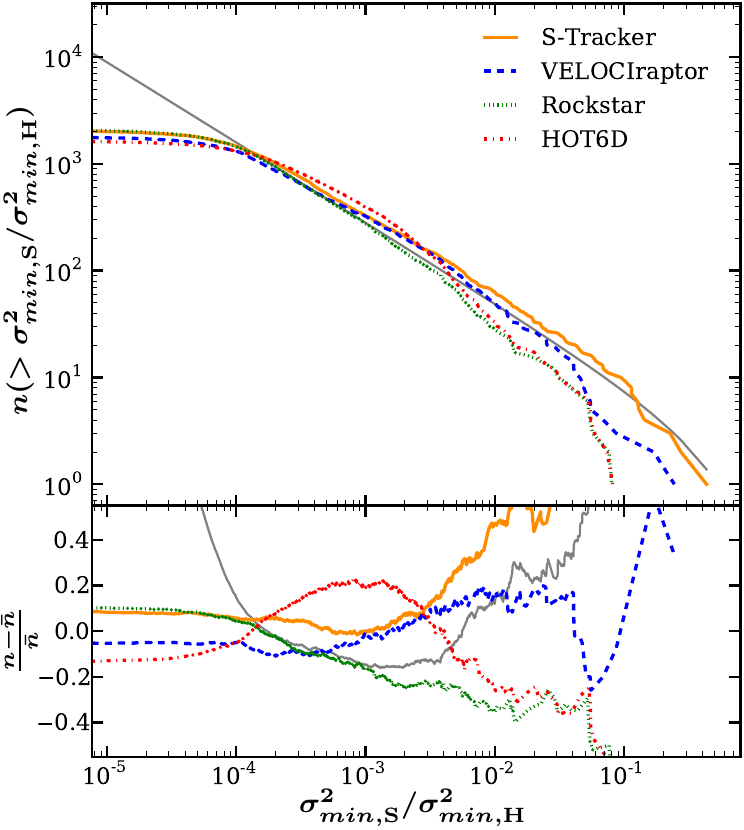}
    \caption{The cumulative velocity dispersion distribution of substructures (top panel) along with the fractional residuals relative to the mean from all the algorithms (bottom panel). We also plot a power-law prediction with $\alpha_\sigma=0.78$. Line and colour scheme are the same as in \Figref{fig:massdistrib}.}
    \label{fig:veldispdistrib}
\end{figure}

Now we turn our attention to the prediction for the dispersion distribution from our cosmological simulation. We remove cloudy tidal debris, which has begun to phase mix with the background and does not a physically meaningful velocity dispersion, and plot the cumulative distribution in \Figref{fig:veldispdistrib} for all finders. 
This distribution like that of the mass function can be characterised by a power-law, specifically the differential distribution is
\begin{align}
  \frac{dn\left(\sigma^2_{min,{\rm S}}/\sigma^2_{min,{\rm H}}\right)}{d\left(\sigma^2_{min,{\rm S}}/\sigma^2_{min,{\rm H}}\right)} \propto \left(\sigma^2_{min,{\rm S}}/\sigma^2_{min,{\rm H}}\right)^{-\alpha_{\sigma}},
\end{align}
where $\sigma^2_{min,{\rm H}}$ is the host's haloes central velocity dispersion. Based on the relation seen in \Figref{fig:massveldispersion} and the power-law of the mass function $\alpha\approx0.9$, we expect $\alpha_\sigma\approx0.78$. This prediction is in good agreement with the data. However, the scatter between the individual finders is quite large, of the order of $20\%$, with large differences at the high dispersion (or mass) end, similar to what is seen in \Figref{fig:massdistrib}. 

\subsection{Radial distribution}\label{sec:radfunc}
\begin{figure}
    \centering
    \includegraphics[width=0.45\textwidth]{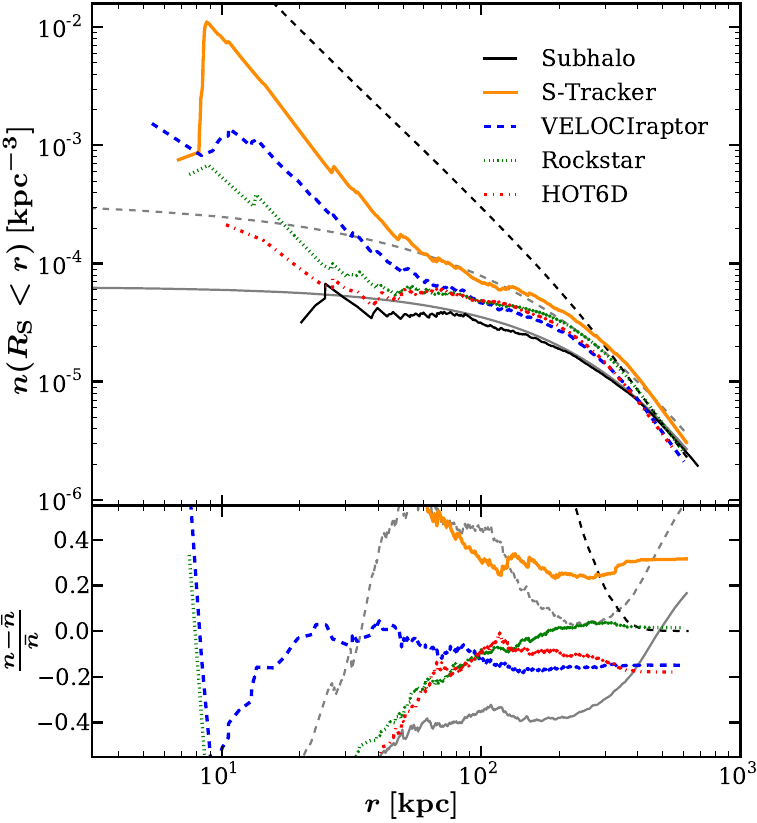}
    \caption{The number density of substructures with centre-of-masses lying within a sphere of radius $R_{\rm S}<r$ for each algorithm (top panel) and the fractional residuals from the mean (bottom panel). We show in the top panel the {\em subhalo} distribution (solid black) and an Einasto fit to this distribution (solid gray), and a fit from \protect\cite{springel2008} based on the highest resolution A-1 halo with $\alpha_{\rm Ein}=0.678$ and $r_{-2}=199$~kpc normalised to the  mean substructure number density at $r_{-2}$ (dashed gray). Line and colour scheme are the same as in \Figref{fig:massdistrib}. For reference, we also show the halo's average density, $\bar\rho(r)\equiv M(r)/(4\pi r^3/3)$, normalized to the mean substructure number density at the edge of the halo (dashed black line).}
    \label{fig:raddistrib}
\end{figure}

Bound subhaloes are more likely to reside in the outskirts of a halo and, although the number density does increase towards the centre, it does so more slowly than the underlying dark matter density \cite[e.g.][]{ghigna1998,gill2004,gao2004,diemand2004,diemand2007,springel2008}. This radial distribution does not depend strongly on the algorithm used to identify the subhalo \cite[cf Fig. 7 in][]{onions2012}\footnote{The most pronounced differences between phase-space finders and configuration-space finders occur in the central region where configuration-space finders, such as {\sc SUBFIND} or {\sc AHF}, will tend to miss subhaloes due to the low density contrast between the substructure and the background \cite[][]{muldrew2011}}. \cite{springel2008} found that the differential number density of subhaloes is well characterised by an Einasto profile, that is 
\begin{align}
  \frac{dn(r)}{dr}=A\exp\left\{(-2/\alpha_{\rm Ein})\left[(r/r_{-2})^{\alpha_{\rm Ein}}-1\right]\right\},
\end{align}
where $\alpha_{\rm Ein}$ is the shape parameter and $r_{-2}$ is the scale radius. We examine the radial {\em substructure} distribution via the number density in a spherical volume $n(R_{\rm S}<r)\equiv(r^3/3)^{-1}\int r^2(dn(r)/dr)dr$, and the differences with respect to the {\em subhalo} distribution in \Figref{fig:raddistrib}. 

\par
This figure shows that the {\em substructure} number density within a radius $r$ decreases monotonically with increasing radius but does so at a slower rate than the {\em subhalo} distribution. The most pronounced difference between the {\em substructure} and {\em subhalo} distribution occurs within the central $\sim50$~kpc, that is within $\sim1-2$ times the radius at which the maximum circular velocity is reached. The subhalo distribution levels off, whereas the substructure distribution continues to increase and has a power-law form. The algorithms do not reproduce the same power-law index, \hst\ having the steepest slope and greatest number density followed by \velociraptor, \rockstar, and \hot6d. Outside the central region, all the algorithms have the same shape though the normalisation differs by $\sim20\%$ as shown by the fact that the residuals at $r>100$~kpc are effectively flat. This outer region does not the same logarithmic slope as the {\em subhalo} number density. The large offset between \hst\ and the other algorithms can be account for by the presence of disrupted substructures with cloud-like morphologies that occupy the same phase-space volume as another bound subhalo in the \hst\ catalogue (see discussion on \Figref{fig:merit} and \Secref{sec:veldispfunc}). 

\par
\begin{figure}
    \centering
    \includegraphics[width=0.45\textwidth]{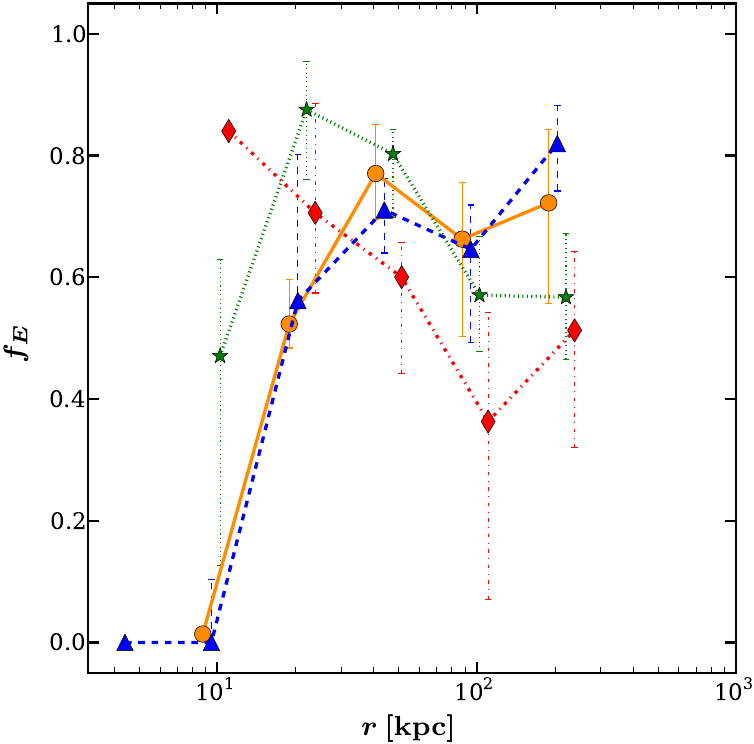}
    \caption{Median and quantiles of the bound mass fraction of substructures as a function of radius. Line and colour scheme is the same as in \Figref{fig:massdistrib}, and markers for \hst, \velociraptor, \rockstar, \hot6d\ are circles, triangles, stars, and diamonds respectively.}
    \label{fig:radfracEdistrib}
\end{figure}

The significant spike in the {\em substructure} number density relative to the {\em subhalo} distribution in the central region is due to the increasing tidal disruption rate of substructures. Figure \ref{fig:radfracEdistrib} clearly shows that the fraction of bound mass in a substructure relative to the mass that is dynamically associated with it, $\fracE$, jumps from about zero to $\sim0.7$ outside a radius of $50$~kpc for \hst, the catalogue containing no false positives, and  \velociraptor. Both of these algorithms have a pronounced power-law component in the central $50$~kpc. \rockstar\ also has a jump in $\fracE$ going from the central most bin to the next but the change is not as pronounced as is the slope of the central power-law component in \Figref{fig:raddistrib}. \hot6d\ is the only algorithm which displays counter-intuitive distribution, with $\fracE$ decreasing with increasing radius, indicating that the algorithm in its current state becomes progressively worse at identifying tidal debris in the high density regions of a halo. 

\par
The agreement between \hst\ and \velociraptor\ suggests that \velociraptor\ has the highest efficiency in disentangling substructure in the dense central regions of a halo. However, the trends in \Figref{fig:raddistrib} are not necessarily a result of one algorithm being more conservative than another, the sensitivity of a finder to noise also plays a role. \velociraptor\ probably performs the best here because it is by construction maximally sensitive to the velocity component, whereas \rockstar\ and \hot6d\ always attempt to use some spatial information, which has a lower signal-to-noise.

\section{Summary} \label{sec:discussion}
We set out to examine the full substructure distribution in a dark matter halo and analyse how well (sub)structure finders that use phase-space information compare with one that also uses temporal information. Recall that {\em substructures} includes {\em subhaloes}, which are a completely self-bound, along with unbound tidal debris. We find excellent agreement between finders regarding pure subhaloes and subhaloes with moderate tidal features, even down to substructures composed of a few tens of particles. Most importantly, all the non-tracking substructure finders are capable of identifying completely unbound tidal debris with a purity of $\sim40\%$. Few tidal debris substructures are missed, though all three non-tracking codes appear to identify spurious tidal debris substructures, comprising $\sim25\%$ of all the tidal debris substructures found, with \hot6d\ having highest fraction of $38\%$ possibly due to the noise in the local phase-space metric used by this finder.

\par
The algorithms also give similar {\em substructure} distributions. First, all algorithms show that, not surprisingly, most subhaloes have pronounced tidal features. The fraction of the host halo's mass in substructures is significantly higher that bound in subhaloes, increasing by almost a factor of two from $\approx10\%$ to $\approx18\%$, with many of these substructures being completely disrupted.  

\par
The mass function does not show any significant variations in shape depending on the algorithm used. The {\em substructure} mass function appears to be slightly steeper than the {\em subhalo} mass function, though at the resolution used in this study it is not possible to make any conclusive statements. However, combined with the results of \cite{maciejewski2011}, we argue that tidal disruption and mergers flattens the distribution. The slope of the substructure distribution suggests that the halo is composed of dynamically distinct substructures all the way down to the bottom of the Cold Dark Matter hierarchy.

\par
The mass function, though often studied, does not lend itself to direct comparison with observations, thus we turn to the velocity dispersion, which is a good proxy for mass and is an observable quantity. The velocity dispersion distribution recovered by the finders are similar, differing by $\approx10-20\%$. Amongst the non-tracking codes, \velociraptor's distribution has the smallest difference with prediction based on the substructure mass function and \hst's distribution.

\par 
The inclusion of tidal debris significantly alters the radial {\em substructure} distribution when compared the {\em subhalo} one. Outside the radius at which the halo's rotation curve peaks, the substructure number density is similar to Einasto profile describing the subhalo number density. Inside this radius the number density is dominated by an additional power-law component. Though all the finders have similar shapes for the radial distribution outside the centre, the slope of the power-law component differs greatly. The tracking code displays the steepest slope and the highest central number density, followed by \velociraptor, \rockstar, and \hot6d. This drastic change in the radial distribution is mirrored by the significant drop in the self-bound mass of substructures. Subhaloes are efficiently tidally stripped in this region. This result does not contradict the findings of \cite{onions2012}, who found that phase-space finders did not fair significantly better than configuration-space finders at identifying subhaloes in the central regions. The ``negligible'' improvement observed is due to the fact that though there are substructures present, effectively none are self-bound {\em at the resolution studied}. With improving resolution and the cumulative effect of missing one subhalo in the centre will lead to completely different two-point correlation functions, making this bias quite significant in the end.

\section{Discussion \& Conclusion}\label{sec:conclusion} 
In this paper, we set out to determine how well substructures finders can recover tidal debris {\em without} the need to track particles. This study is a follow-up to the {\it Haloes Gone Mad} \citep{knebe2011}, {\it Subhaloes Going Notts} \citep{onions2012}, and {\it Galaxies Going Mad} \citep{knebe2012a} comparison projects. The motivation for this comparison study is two-fold. The hierarchical paradigm of structure formation predicts that dark matter haloes are littered with the remnants of disrupted subhaloes. Disrupted dark matter subhaloes may leave signatures that could be detected by direct dark matter detectors as these experiments are sensitive to the local velocity distribution. Though the {\em subhalo} distribution has been well studied, the full {\em substructure} distribution has not. Additionally, upcoming large simulations will include built-in finders and may not store enough snapshot to accurately track tidally disrupted objects, thus requiring finders capable of detecting unbound substructures. 

\par
Second, observations demonstrate that galaxies do not have quiet lives, as shown by disrupted satellite galaxies such as the Sagittarius stream \cite[e.g.][]{ibata2001b,belokurov2006}. Upcoming missions like GAIA will find even more examples of murdered satellites. Yet the study of tidal debris in regards to Galactic Archaeology and Near Field Cosmology is still in its infancy. Searching for tidal debris in the GAIA data set requires the development of new tools as one cannot simply follow the orbital evolution of ``stars'' (particles) to identify tidal debris as is often done with numerical simulations. Hence an understanding of how well substructure finders which solely use phase-space information is sorely needed as these tools could be applied directly to observational data.

\par
We conclude that several currently available substructure finders can correctly identify tidal debris and the tidal features associated with subhaloes in fully cosmological simulations without the need for temporal information. Based on the cross-matching of individual substructures and a comparison of the observed substructure distributions, we argue the most promising non-tracking methods tested here are \velociraptor\ and \rockstar. We find all non-tracking algorithms identify tidal features associated with tidally disrupted subhaloes with a high degree of purity, $\sim85\%$. The purity of {\em completely unbound} substructures is worse at $\sim40\%$ and there also is a non-negligible fraction of spurious tidal debris substructures. Though our tests conclusively show that these algorithms work reasonably well given the difficult challenge of identifying completely unbound dark matter substructures, there are several shortcomings we do not address. The complex and diffuse nature of tidal debris means that non-tracking codes may split a single structure into several groups, biasing the number distribution. One possible method of accurately merging patches of a stream would be to estimate the elliptical orbit of each patch and determine the likelihood that two or more ``streams'' have the same orbit and similar velocity dispersions. Finally, the tracking code will continue to follow dark matter substructures that appear to have effectively phase-mixed with the background that are missed by the non-tracking finders. This may not be a serious limitation as many of these objects were completely disrupted more than several Gyrs ago. Whether these objects should be considered distinct substructures or not and determining the time-scale over which a substructure completely phase-mixes is an open question and is beyond the scope of this study.

\par
We have also shown that a significant fraction of a halo's mass is in tidal debris, similar to the amount of mass in subhaloes. We have only scratched the surface of the information that can be retrieved from tidal debris and the tidal features surrounding subhaloes in this paper. For example, semi-analytic models of galaxy formation  (SAM) use halo merger trees and the current mass of a (sub)halo to determine the galaxy that resides in said (sub)halo \cite[e.g.][]{benson2011,guo2011}. Using the algorithms tested here would allow these models to include the dynamical state of the substructure to alter the galaxy residing in the substructure by accounting for for stellar and gas mass loss, and study the formation of the main galaxy's stellar halo. Currently, the few theoretical studies examining the formation of our Galaxy's stellar halo use semi-analytic models to ``paint'' stars on a dark matter only simulation and evolve this simulation forward in time \cite[e.g][]{cooper2010,helmi2011}, which is significantly more computationally intensive that just using a SAM and certainly prone to limitations as shown by \cite{libeskind2011}.

\par
Another application for these algorithms is in the study of tidally disrupted satellites like the Sagittarius stream and the Orphan stream. The properties of these stellar streams have been used to infer the potential of our Galaxy \cite[e.g.][]{penarrubia2005,penarrubia2006,law2010,koposov2010,deg2012} and could be used to test gravity \cite[e.g.][]{penarrubia2012} or search for unseen dark matter subhaloes \cite[e.g.][]{yoon2011}. Typically, these studies examine the velocities and positions of stars in the stream and compare them to simple numerical simulations or models which assume a constant or slowly evolving host potential and neglect the interactions between satellites or substructure in the satellite itself. In cases such as the GD-1 stream, satellite-satellite interactions are probably negligible since the stream is extremely narrow \cite[e.g.][]{koposov2010}, though in general this will not be the case. Streams are not the only tidal debris which can be used to study either models of the Milky Way or different forms of gravity. \cite{sanderson2012} showed that the radial density profiles of shell-like tidal debris in simple toy models can be used to infer the radial gravitational potential at the radial position of the shell. A more complete understanding of the distribution of tidal debris and the effects of satellite interactions on them from cosmological simulations would be invaluable to these studies. 

\par
We conclude that these algorithms will open up a new window on galaxy formation by allowing us to study a halo's merger history in more detail using the complete substructure distribution and directly search observational data from future missions such as GAIA for the remnants of tidally disrupted satellite galaxies.

\section*{Acknowledgements}
The work in this paper was initiated at the Subhaloes going Notts workshop in Dovedale, UK, which was funded by the European Commissions Framework Programme 7, through the Marie Curie Initial Training Network Cosmo-Comp (PITN-GA-2009-238356). We basically thank all the participants of that meeting for all the stimulating discussions and the great time in general and thank the anonymous referee for useful comments on the manuscript.

\par
PJE and JXH acknowledges financial support from the Chinese Academy of Sciences (CAS), from NSFC grants (No. 11121062, 10878001,11033006), by the CAS/SAFEA International Partnership Program for Creative Research Teams (KJCX2-YW-T23). JXH and HL acknowledge a fellowship from the European Commissions Framework Programme 7, through the Marie Curie Initial Training Network CosmoComp (PITN-GA-2009-238356). YA is supported by the {\it Spanish Ministerio de Ciencia e Innovaci\'on} (MICINN) through  the Ramon y Cajal programme, as well as grant AYA 2010-21887-C04-03. also AK is supported by MICINN through the Ramon y Cajal programme as well as the grants AYA 2009-13875-C03-02, AYA2009-12792-C03-03, CSD2009-00064, and CAM S2009/ESP-1496 and the {\it Ministerio de Econom\'ia y Competitividad} (MINECO) through grant AYA2012-31101. He further thanks Serge Gainsbourg for the roller girl.

\pdfbookmark[1]{References}{sec:ref}
\bibliographystyle{mn2e}
\bibliography{SgN.bbl}

\end{document}